\documentclass[%
superscriptaddress,
amsmath,amssymb,
aps,
prx,
longbibliography,
11pt
]{revtex4-2}

\usepackage{amssymb}
\usepackage{amsthm}
\usepackage{graphicx}
\usepackage{amsmath}
\usepackage{amsfonts}
\usepackage{color}
\usepackage{centernot}
\usepackage{mathtools}
\usepackage{stmaryrd}

\begin{document}
\title{Hidden scale invariance of turbulence in a shell model: \\ from forcing to dissipation scales}
\author{Alexei A. Mailybaev} 
\affiliation{Instituto de Matem\'atica Pura e Aplicada -- IMPA, Rio de Janeiro, Brazil}
\email{alexei@impa.br}

\begin{abstract}
Intermittency is one of central obstacles for understanding small-scale dynamics in the fully developed hydrodynamic turbulence. The modern approach is largely based on the multifractal theory of Parisi and Frisch which is, however, phenomenological. It was shown recently that the intermittency can be related to the hidden scale invariance. The latter is a new statistical scaling symmetry unbroken in a rescaled (projected) formulation of equations of motion. In the present work, we consider a shell model of turbulence and describe how the hidden symmetry manifests itself through all scales, both in the inertial interval and in the transition to forcing and dissipation ranges. In the inertial interval, we derive anomalous scaling laws from the hidden symmetry. Then, we show how a complicated form of the dissipation range is controlled by intermittent rescaled Reynolds numbers within a large range of dissipation scales. This dissipative intermittency can be removed by using a special class of dissipation models. For such models, the hidden scale invariance is restored both in the inertial interval and the dissipation range. Overall, the presented approach deduces the multifractal theory and some of its basic conclusions from the hidden scaling symmetry of equations of motion.
\end{abstract}

\maketitle

\section{Introduction} 

In stationary statistics of fully developed hydrodynamic turbulence, one traditionally distinguishes the forcing range of large scales, the dissipation range of very small scales, and the inertial interval in between~\cite{frisch1999turbulence}. At scales of the inertial interval, both forcing and viscous forces are negligible. Description of these regions relies on the scale invariance of the underlying Navier--Stokes equations. In the absence of (or far from) physical boundaries these symmetries are formulated as
    \begin{equation}
	t,\ \mathbf{x},\ \mathbf{u},\ \nu \ \ \mapsto \ \ 
	\lambda^{1-h}t,\ \lambda \mathbf{x},\ \lambda^h \mathbf{u},\ \lambda^{1+h}\nu,
    \label{eqI1}
    \end{equation}
where $\mathbf{u}(\mathbf{x},t)$ is a velocity field and $\nu$ is a kinematic viscosity. Relation (\ref{eqI1}) defines a family of space-time scaling symmetries depending on two real parameters, $\lambda > 0$ and $h \in \mathbb{R}$. One of major obstacles for the theory of turbulence is that all scaling symmetries (\ref{eqI1}) are broken in the stationary statistics, contrary to the initial self-similarity hypothesis of Kolmogorov's 1941 (K41) theory relying on $h = 1/3$~\cite{kolmogorov1941local,frisch1999turbulence}.

The observed statistics in the inertial interval shows that moments of velocity fluctuations $\delta u_\ell$ depend on a scale $\ell$ in the form of power laws $\langle \delta u_\ell^p\rangle \propto \ell^{\zeta_p}$. Scaling exponents $\zeta_p$ in these relations depend nonlinearly on $p$, which is the manifestation of  small-scale intermittency. Such (so-called anomalous) scaling also affects a large part of the dissipation range~\cite{frisch1993prediction}. The observed intermittency is successfully described by the Parisi--Frisch multifractal theory~\cite{frisch1985singularity,frisch1999turbulence} associating different fractal dimensions to different exponents $h$ in Eq.~(\ref{eqI1}). This theory, however, is phenomenological, i.e., it does not follow from equations of motion. 

Till now, three-dimensional (3D) incompressible Navier--Stokes system does not allow numerical simulations at very large Reynolds numbers; see e.g.~\cite{iyer2021oscillations}. For this reason, much attention is paid to simplified (toy) models as a playground for testing theoretical ideas. Shell models~\cite{gledzer1973system,ohkitani1989temporal,biferale2003shell} is a such class of models, which successfully describe intermittent properties of turbulence; we refer to~\cite{benzi1993intermittency,l2000analytic,benzi2003intermittency,eyink2003gibbsian,PhysRevX.11.021063} for some related studies in this direction. In the present work, we focus on one of most popular shell models called the Sabra model~\cite{l1998improved}.

It was shown recently that the anomalous scaling of structure functions can be generated by Perron--Frobenius eigenmodes of the hidden scaling symmetry~\cite{mailybaev2020hidden}. This new symmetry refers to equations of ideal fluid dynamics written for dynamically rescaled velocities and time. Geometrically, the rescaling procedure is a projection in phase space, which enables the emergence of new symmetries. Unlike the broken scaling symmetries (\ref{eqI1}), there is numerical evidence that the hidden scale invariance is restored in the inertial interval both for the Navier--Stokes system~\cite{mailybaev2022hidden} and shell models~\cite{mailybaev2021hidden}. The emerging hidden-symmetry formalism naturally unifies the self-similarity ideas going back to Kolmogorov~\cite{kolmogorov1962refinement,benzi1993intermittency,eyink2003gibbsian} with the multifractality of Parisi and Frisch~\cite{frisch1985singularity,mailybaev2022shell}. 

The present work reports on further developments in this direction. First, we reveal some extra details on scaling laws, including the relation of hidden symmetry with the anomalous dissipation and multifractal spectrum. Our central goal, however, is the analysis of transition between the inertial interval and the dissipation (or forcing) range. We argue that this transition is controlled by the hidden-symmetric state and its stability. For example, a transition to the forcing range is governed  by an exponentially decaying mode. The situation is very different in the dissipation range, because viscous terms in the rescaled formulation are intermittent. This leads to a gradual breaking of the hidden symmetry within a large range of scales. Lastly, we show that the intermittent dissipation is an artifact of a specific (viscous) dissipation terms in the equations of motion. We introduce a class of models with a viscous cutoff, in which the hidden symmetry can be extended to all small scales, including the dissipative ones. As an application of this symmetry, we derive a functional form of structure functions valid at all small scales, both in the inertial interval and dissipation range. 

Our general conclusion, is that the concept of hidden symmetry provides a firm theoretical basis for the small-scale analysis of fully developed turbulence. This approach reformulates (and validates in their new form) the original Kolmogorov's ideas, in which the $h = 1/3$ scaling symmetry is replaced by the hidden scale invariance.

We start with a description of the Sabra shell model and its rescaled version in Section~\ref{sec2}. Section~\ref{sec_intII}  defines and verifies the hidden scaling symmetry in the inertial interval, and studies its consequences for structure functions, energy flux and large deviations. Section~\ref{sec_visc} uses the hidden symmetry for studying a transition from the forcing and dissipation ranges to the inertial interval. Section~\ref{sec_diss} presents a class of viscous-cutoff shell models, which possess the extended form of hidden scale invariance at dissipation scales. We conclude by discussing the obtained results and their applicability to the Navier--Stokes system. Some technical derivations are gathered in the Appendix.

\section{Shell model and its rescaled representation} \label{sec2}

Shell models of turbulence mimic the Navier--Stokes flow using a geometric sequence of spatial scales $\ell_n = 2^{-n}\ell_0$, where $\ell_0$ is an integral scale and $n$ an integer shell number. Thus, large scales correspond to shell numbers around zero, while small scales are given by large shell numbers. The associated wavenumbers are defined as $k_n = 1/\ell_n = 2^n k_0$. Velocity fluctuations $\delta u(\ell_n)$ at different scales are represented by complex variables $u_n \in \mathbb{C}$, which are called shell velocities. We denote by $u =  (u_n)_{n \in \mathbb{Z}}$ the full (bi-infinite) state vector. 

\subsection{Shell model equations and scaling symmetries} \label{subsec_SME}

We consider the constant forcing at the integral scale expressed via the boundary conditions 
    \begin{equation}
	u_0(t) \equiv u_0 > 0, \quad u_n(t) \equiv 0 \ \ \textrm{for}\ \ n < 0,
    \label{eq2b}
    \end{equation}
where $u_0$ is a real positive constant.
The Sabra shell model~\cite{l1998improved} is formulated as
    \begin{equation}
    \frac{du_n}{dt} = k_0\mathcal{B}_n[u]-\nu k_n^2u_n, \quad n > 0,
	\label{eq1a}
    \end{equation}
where $\mathcal{B}_n[u]$ is a quadratic form defined as
    \begin{equation}
	\mathcal{B}_n[u] = i2^n \left(2u_{n+2}u_{n+1}^*
	-\frac{u_{n+1}u_{n-1}^*}{2} 
	+\frac{u_{n-1}u_{n-2}}{4}\right),
	\label{eq1Bn}
    \end{equation}
and $\nu \ge 0$ is a viscosity parameter. The quadratic term in Eq.~(\ref{eq1a}) imitates the convective and pressure terms of the Navier--Stokes system.
It is designed such that the shell model possesses two inviscid invariants, the energy $\mathcal{E}[u] = \frac{1}{2}\sum_n |u_n|^2$ and helicity $\mathcal{H}[u] = \sum_n(-1)^nk_n|u_n|^2$, analogous to the invariants in 3D ideal flows~\cite{l1998improved}. 

The dimensionless Reynolds number is defined as $\mathrm{R} = u_0\ell_0/\nu$. 
In this work, we describe the fully developed turbulent state, i.e., the stationary (long-time) statistics for very large Reynolds numbers.
By saying very large we mean that both $\mathrm{R}$ and its logarithm are large. 
We remark that the limit $\mathrm{R} \to \infty$ is often studied as the limit of small viscosity $\nu \to 0$ with fixed $\ell_0$ and $u_0$. 

One can see that Eqs.~(\ref{eq1a})--(\ref{eq1Bn}) are invariant with respect to space-time scalings of the form
    \begin{equation}
    t,\ u_n,\ \nu \ \ \mapsto \ \ 2^{1-h}t,\ 2^h u_{n+1},\ 2^{1+h}\nu.
    \label{eq_S1}
    \end{equation}
Here the exponent $h \in \mathbb{R}$ defines an arbitrary factor $2^{1-h}$ for time scaling, and the shift of shell numbers mimics the space scaling because $\ell_{n+1} = \ell_n/2$. Hence, transformations (\ref{eq_S1}) generate a symmetry group of  space-time scalings (discrete in space and continuous in time), which is analogous to space-time scaling symmetries (\ref{eqI1}) of the Navier--Stokes equations. These symmetries are broken at the integral scale $\ell_0$ by the boundary conditions (\ref{eq2b}).

\subsection{Rescaled velocities and time} \label{subsec_RV}
The theory we develop in this work is based on the hidden scaling symmetry. This symmetry is different from and weaker than symmetries (\ref{eq_S1}), i.e., it can be restored in a statistical sense even when all symmetries (\ref{eq_S1}) are broken. Such a new symmetry emerges when equations of motion are written in terms of rescaled (projected) variables as we describe below.

Let us fix some reference shell number $m \ge 0$. We define a corresponding state-dependent velocity amplitude $\mathcal{A}_m[u]$ and a temporal scale (turn-over time) $\mathcal{T}_m[u]$ as 
    \begin{equation}
	\mathcal{A}_m[u] = \sqrt{\sum_{j \ge 0}{\alpha^j|u_{m-j}|^2}},\quad 
	\mathcal{T}_m[u] = \frac{\ell_m}{\mathcal{A}_m[u]}.
    \label{eq2A}
    \end{equation}
In this expression, the role of the pre-factors $\alpha^j$ is to suppress the contribution from distant scales (much larger than $\ell_m$) and also ensure that the amplitude is strictly positive. As we show in Section~\ref{subsec_ldt}, for all our purposes it is enough to choose $0 < \alpha < 0.4$. We use $\alpha = 1/8$ in the numerical simulations.

We now normalize all variables with respect to the reference shell $m$; see Eq.~(\ref{eq2A}). This yields the new rescaled velocities $U_N^{(m)}$ as functions of the intrinsic time $\tau^{(m)}$ defined implicitly as
    \begin{equation}
    U_N^{(m)} = \frac{u_{m+N}}{\mathcal{A}_{m}[u]},\quad
    d\tau^{(m)} = \frac{dt}{\mathcal{T}_{m}[u]},
    \label{eq2}
    \end{equation}
with the initial time $\tau^{(m)} = 0$ corresponding to $t = 0$. We denote by $U^{(m)} = \big(U_N^{(m)}\big)_{N \in \mathbb{Z}}$ the full (bi-infinite) rescaled state. 

For $m = 0$ and boundary conditions (\ref{eq2b}), we find $\mathcal{A}_{m}[u] = u_0$ and the rescaled variables reduce to the usual dimensionless form $U_N^{(0)} = u_N/u_0$ and $\tau^{(0)} = t u_0/\ell_0$. 
Expressions (\ref{eq2A})--(\ref{eq2}) considered for $m+1$ yield
    \begin{equation}
	    \mathcal{A}_{m+1}[u]
	    = \sqrt{|u_{m+1}|^2+\sum_{j \ge 1}{\alpha^{j}|u_{m+1-j}|^2}} 
	    = \sqrt{|u_{m+1}|^2+\alpha \mathcal{A}^2_m[u]}
	    = \mathcal{A}_m[u]\sqrt{\alpha+\big|U_1^{(m)}\big|^2}
    \label{eqA1_Um}
    \end{equation}
and, as a consequence, the expressions 
    \begin{equation}
	U_N^{(m+1)} = \frac{U_{N+1}^{(m)}}{\sqrt{\alpha+\big|U_1^{(m)}\big|^2}},\quad 
	d\tau^{(m+1)} = 2\sqrt{\alpha+\big|U_1^{(m)}\big|^2}\, d\tau^{(m)}
	\label{eqI_HS_m}
    \end{equation}
relating the rescaled variables for different reference shells.
 
One can check using (\ref{eq2A})--(\ref{eq2}) that the rescaled velocities satisfy the identity
    \begin{equation}
    \mathcal{A}_0[U^{(m)}] = \sqrt{\sum_{j \ge 0} \alpha^j \big|U_{-j}^{(m)}\big|^2} = 1.
    \label{eqAp_3b}
    \end{equation}   
Hence, transformation (\ref{eq2}) for velocity variables can be seen as a projection onto the hypersurface (\ref{eqAp_3b}); see \cite{mailybaev2020hidden} for a general theory of such projections, which are related to the time scaling. In particular, multiplying original velocities $u$ by any positive factor leaves the rescaled velocities $U^{(m)}$ intact. Thus, the rescaling is not invertible: one cannot express $u$ in terms of $U^{(m)}$ alone.

We remark that a specific choice of $\mathcal{A}_m[u]$ in (\ref{eq2A}) is not particularly important. In fact, one can write an equivalent formulation of hidden symmetry for a large class of positive, homogeneous and scale invariant expressions for $\mathcal{A}_m[u]$; see~\cite{mailybaev2020hidden} for more details. 

\subsection{Rescaled equations}\label{subsec_RE}

Performing the transformation (\ref{eq2}), the shell model equations (\ref{eq1a}) take the form (see \cite{mailybaev2022shell,juliathesis} and Appendix~\ref{secA1} for derivations)
    \begin{equation}
    \begin{array}{rcl}
    \displaystyle
    \frac{dU_N^{(m)}}{d\tau^{(m)}} 
    & = & \displaystyle
    \mathcal{B}_N[U^{(m)}]
         -U_N^{(m)}\sum_{j = 0}^{m-1}\alpha^j\mathrm{Re}\left(U_{-j}^{(m)*}\mathcal{B}_{-j}[U^{(m)}]\right) \\[15pt]
    && \displaystyle
	-\, \frac{U_N}{\mathrm{R}_m[u]} \bigg(4^N-\sum_{j = 0}^{m-1} \alpha^j4^{-j}|U_{-j}|^2 \bigg), \quad N > -m.
	\end{array}
	\label{eq1Ra}
    \end{equation}
Here $\mathrm{Re}(\cdot)$ denotes the real part, quadratic terms $\mathcal{B}_N[U]$ are given by Eq.~(\ref{eq1Bn}), and we introduced the local Reynolds number as
    \begin{equation}
	\mathrm{R}_m[u] = \frac{\mathcal{A}_m[u]\ell_m}{\nu}.
	\label{eq1Rb}
    \end{equation}
Analogous transformation of boundary conditions (\ref{eq2b}) yields (see Appendix~\ref{secA1}) 
   \begin{equation}
	U_{-m}^{(m)} = \sqrt{\alpha^{-m}-\sum_{j = 0}^{m-1} \alpha^{j-m}\big|U_{-j}^{(m)}\big|^2},\quad
	U_{N}^{(m)} = 0 \ \ \textrm{for} \ \ N < -m.
	\label{eq1R_bc}
    \end{equation}

Equations (\ref{eq1Ra})--(\ref{eq1R_bc}) define a rescaled system for the velocities $U_N^{(m)}$ as functions of $\tau^{(m)}$. Notice that $\mathrm{R}_m[u]$ is not expressed in terms of $U^{(m)}$, i.e., viscous terms of the rescaled system are not expressed in terms of rescaled variables.

\section{Intermittency in the inertial interval}
\label{sec_intII}

In a classical description of fully developed turbulence, the forcing is limited to large scales of order $\ell_0$ (the forcing range), while the viscous effects become considerable only at very small scales (the dissipation range)~\cite{frisch1999turbulence}. The scales in between form the so-called inertial interval, where both forcing and viscous effects can be neglected. In this section, we establish the hidden scale invariance in the inertial interval, and show that the phenomenon of intermittency is a consequence of this new symmetry. Thus, we naturally identify the inertial interval with scales, at which the hidden symmetry is restored in a statistical sense. Then, boundary conditions and viscous terms break the hidden symmetry at scales of the forcing and dissipation ranges. The precise extent of the inertial interval is derived later in Section~\ref{sec_visc} from the analysis of hidden-symmetry breaking. We anticipate this result here and define the scales $\ell_n$ of inertial interval as 
    \begin{equation}
    \mathrm{R}^{-\frac{1}{1+h_{\max}}} 
    \ll \frac{\ell_n}{\ell_0} \ll 1,
	\label{eq_IRdef}
    \end{equation}
where the exponent $1/(1+h_{\max}) \approx 0.58$ corresponds to the maximum H\"older exponent $h_{\max}$ of a hidden-symmetric state.

\subsection{Hidden scale invariance}
\label{subsec_HS}

In the present section, we assume that all shell numbers under consideration belong to the inertial interval. Therefore, we ignore both viscous and forcing (boundary) effects in system (\ref{eq2b})--(\ref{eq1Bn}). We refer to the resulting system of equations as the ideal shell model, which takes the form
    \begin{equation}
    \frac{du_n}{dt} = k_0 \mathcal{B}_n[u], \quad n \in \mathbb{Z}.
	\label{eq1Euler}
    \end{equation}
This system imitates the Euler equations for ideal fluid. Its scaling symmetries have the form
    \begin{equation}
    t,\ u_n \mapsto 2^{1-h}t,\ 2^h u_{n+1}
    \label{eq_S1ideal}
    \end{equation}
for any $h \in \mathbb{R}$, which follow from Eq.~(\ref{eq_S1}) ignoring the relation for viscosity.

Let us transform system (\ref{eq1Euler}) to equations for rescaled variables (\ref{eq2}). This yields
    \begin{equation}
    \frac{dU_N}{d\tau} = \mathcal{B}_N[U]
         -U_N\sum_{j \ge 0} \alpha^{j}\mathrm{Re}\left(U_{-j}^{*}\mathcal{B}_{-j}[U]\right), \quad N \in \mathbb{Z},
	\label{eqI_R}
    \end{equation}
which follows from Eq.~(\ref{eq1Ra}) after dropping the viscous terms and the boundary (large-scale) limit in the sums.
Here we also (temporarily) dropped the superscript $(m)$, thereby stressing that the rescaled ideal system (\ref{eqI_R}) does not depend on a choice of the reference shell. 
Hence, this system has a symmetry corresponding to a shift of the reference shell:
    \begin{equation}
	m \mapsto m+1.
	\label{eqI_HSi}
    \end{equation}
Explicit form of this symmetry follows from relations (\ref{eqI_HS_m}) as 
    \begin{equation}
	U,\ \tau \ \mapsto \ \hat{U},\ \hat{\tau},
	\label{eqI_HS}
    \end{equation}
where the new rescaled state $\hat{U} = \big(\hat{U}_N\big)_{N \in \mathbb{Z}}$ and time $\hat{\tau}$ are defined as 
    \begin{equation}
	\hat{U}_N = \frac{U_{N+1}}{\sqrt{\alpha+|U_1|^2}},\quad d\hat{\tau} = 2\sqrt{\alpha+|U_1|^2}\, d\tau.
	\label{eqI_HSexp}
    \end{equation}
In Appendix \ref{secA_2} we show explicitly that system (\ref{eqI_R}) is invariant with respect to the transformation (\ref{eqI_HS})--(\ref{eqI_HSexp}). This is what we call the \textit{hidden scaling symmetry}. 

The important property of transformation (\ref{eqI_HS})--(\ref{eqI_HSexp}) is that it defines a statistical symmetry~\cite{mailybaev2020hidden}. 
Statistical properties of the rescaled system are computed with the rescaled time $\tau$, and they can be tested using averaged observables (test functions) $\varphi(U)$ as
    \begin{equation}
	\langle \varphi(U) \rangle_{\tau} = \lim_{T \to \infty}\frac{1}{T} \int_0^T \varphi(U(\tau)) \,d\tau = \int \varphi(U)\, d\mu(U).
	\label{eqI_HSst}
    \end{equation}
The last expression contains  a probability measure $d\mu(U)$ of the statistically stationary state. 
The hidden symmetry transformation (\ref{eqI_HS})--(\ref{eqI_HSexp}) changes the statistics, i.e., transforms the stationary probability measure $\mu \mapsto \hat{\mu}$.
Denoting the state transformation in (\ref{eqI_HS}) as $\hat{U} = G(U)$, the new measure is expressed using the standard dynamical system analysis~\cite{cornfeld2012ergodic} as
    \begin{equation}
        \hat{\mu} = G_\sharp \bar{\mu}, \quad d\bar{\mu}(U) = \frac{\sqrt{\alpha+|U_1|^2}}{\langle \sqrt{\alpha+|U_1|^2}\rangle_{\mu}}\, d\mu(U).
	\label{eqI_HSstM}
    \end{equation}
Here the pushforward in the first expression corresponds to the change of state, while the second expression reflects the change of time; see \cite{mailybaev2020hidden} for more details and precise mathematical formulations.
We say that the hidden symmetry is restored in the statistical sense if the probability measure remains invariant under the hidden symmetry transformation. By construction, such property implies the statistical self-similarity: statistics of the rescaled state $U^{(m)}$ with respect to time $\tau^{(m)}$ does not depend on the reference shell $m$.

Transformation (\ref{eqI_HS})--(\ref{eqI_HSexp}) does not depend on the parameter $h$ and, therefore, it is not equivalent to any of the original scaling symmetries (\ref{eq_S1ideal}). 
In particular, the hidden symmetry may be restored for a statistically stationary state even when all original scaling symmetries (\ref{eq_S1ideal}) are broken. We refer to \cite{mailybaev2021hidden} for a general theory and \cite{mailybaev2021solvable} for an analytic example.
Numerical simulations of~\cite{mailybaev2021hidden,mailybaev2022shell} and our results below strongly support the conjecture that the hidden symmetry is restored for the statistics of rescaled variables within the inertial interval. From now we assume this property and analyze its consequences for the turbulent dynamics.

\subsection{Universality of Kolmogorov multipliers}
Let us return to use the superscript $(m)$ for the rescaled variables. In this subsection, we consider the so-called Kolmogorov multipliers inspired by Kolmogorov's ideas of 1962~\cite{kolmogorov1962refinement}. For a shell model, these multipliers were defined as the ratios $|u_n/u_{n-1}|$ in~\cite{benzi1993intermittency,eyink2003gibbsian}. Numerical studies reported in these works suggested that single-time statistics of multipliers is independent of the shell number $n$ in the inertial interval, and this universality motivated the first formulation of hidden symmetry~\cite{mailybaev2021hidden}. Using definition (\ref{eq2}), multipliers are expressed in terms of the rescaled variables as 
    \begin{equation}
	\label{eq_Km1}
	\left|\frac{u_n}{u_{n-1}}\right| 
	= \left| \frac{U_{N}^{(m)}}{U_{N-1}^{(m)}} \right|,\quad n = m+N.
    \end{equation}
As a consequence of the statistical hidden symmetry, the statistics of ratios $|U_{N}^{(m)}/U_{N-1}^{(m)}|$ considered as functions of $\tau^{(m)}$ do not depend on the reference shell $m$. Using this self-similarity property, one can show the universality of single-time (but not multi-time) statistics of multipliers as functions of the original time $t$~\cite{mailybaev2022hidden}.

For our purposes, we introduce a generalized version of Kolmogorov multipliers as ratios of velocity amplitudes: $\mathcal{A}_{n}[u]/\mathcal{A}_{n-1}[u]$.
This definition has several advantages compared to Eq.~(\ref{eq_Km1}), e.g., avoiding pathologies caused by a vanishing denominator. Taking $n = m+N$, we express the generalized multiplier in terms of rescaled velocities (\ref{eq2}) as (see Appendix~\ref{subsec_GMult})
    \begin{equation}
	\frac{\mathcal{A}_{n}[u]}{\mathcal{A}_{n-1}[u]} = \mathcal{X}_N[U^{(m)}],\quad
	n = m+N,
	\label{eqKm}
    \end{equation}
where the functions $\mathcal{X}_N[U]$ are defined as 
    \begin{equation}
	\mathcal{X}_N[U]
	= \sqrt{\alpha+\frac{\alpha|U_N|^2}{\sum_{j \ge 1}{\alpha^{j}|U_{N-j}|^2}}}.
	\label{eqKm2}
    \end{equation}
Notice that all $\mathcal{X}_N[U] \ge \sqrt{\alpha}$. 

\begin{figure}[t]
\centering
\includegraphics[width=0.85\textwidth]{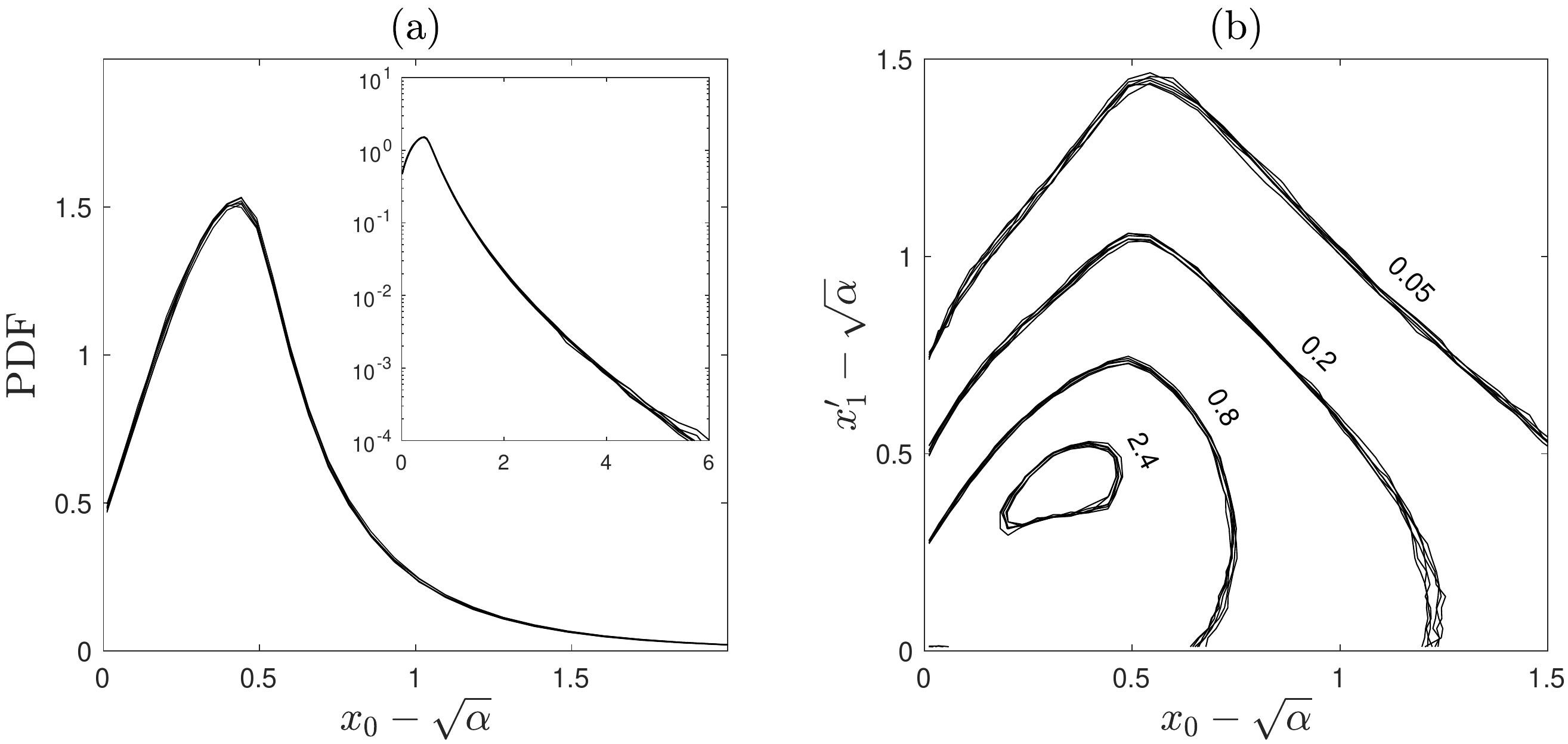}
\caption{(a) Collapse of PDFs for the multipliers $x_0 = \mathcal{X}_0[U^{(m)}]$ computed for the shells $m = 7,\ldots,12$ from the middle of inertial interval; the inset shows the same graphs with a logarithmic vertical scale. (b) Collapse of level curves for the joint two-time PDF of the multipliers $x_0 = \mathcal{X}_0[U^{(m)}(\tau^{(m)})]$ and $x'_1 = \mathcal{X}_1[U^{(m)}(\tau^{(m)}+1)]$ for $m = 7,\ldots,12$. The results use simulations with $\mathrm{R} = 10^{10}$ and $T = 10^4$. For convenience, we subtracted the minimum value $\sqrt{\alpha}$ in multiplier axes.}
\label{fig1}
\end{figure}

We verify the statistical hidden scale invariance of multipliers (\ref{eqKm}) numerically using a long-time simulation of shell model (\ref{eq2b})--(\ref{eq1Bn}) with $\mathrm{R} = 10^{10}$; see Appendix~\ref{subsecA_num} for details of numerical simulations. Figure \ref{fig1}(a) shows an accurate collapse of probability density functions (PDFs) of the multiplier $x_0 = \mathcal{X}_0[U^{(m)}]$, where the shells $m = 7,\ldots,12$ are chosen from the middle of inertial interval (\ref{eq_IRdef}). Figure \ref{fig1}(b) shows a similar collapse for two-time joint PDFs of the multipliers $x_0 = \mathcal{X}_0[U^{(m)}(\tau^{(m)})]$ and $x'_1 = \mathcal{X}_1[U^{(m)}(\tau^{(m)}+1)]$ taken at the rescaled-time interval $\Delta\tau^{(m)} = 1$. We stress that the use of rescaled time $\tau^{(m)}$ is crucial for the universality of multi-time statistics. Such universality does not hold for multipliers as functions of the original time $t$.

\subsection{Structure functions and scaling laws}\label{subsec_SF}

Structure functions are traditional observables for the analysis of intermittency in fully developed turbulence \cite{frisch1999turbulence}. 
For a shell model, structure functions are usually defined as time-averaged velocity moments, $\langle |u_n|^p \rangle_t$ for $p \in \mathbb{R}$. 
These averages, however, diverge for $p \le -2$. In this subsection, we consider the moments of velocity amplitudes $\mathcal{A}_m[u]$ given by Eq.~(\ref{eq2A}), which do not diverge. Other formulations are considered in Section~\ref{subsec_genSF}. 

\begin{figure}[t]
\centering
\includegraphics[width=0.9\textwidth]{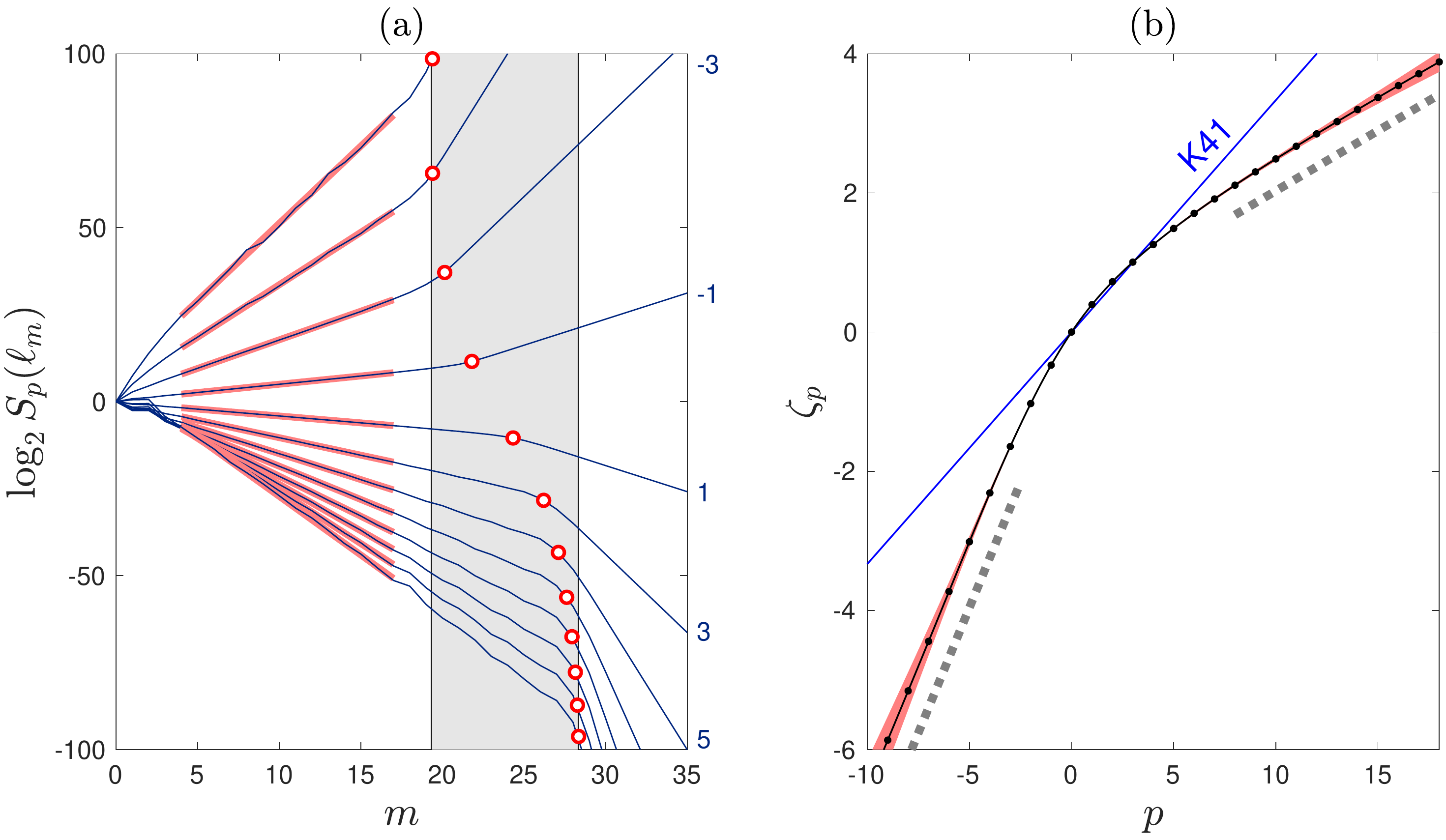}
\caption{(a) Logarithms of structure functions $S_p(\ell_m) = \left\langle \mathcal{A}_m^p[u] \right\rangle_t$ for odd orders $p = -7,-5,\ldots,15$. Bold red lines show the power law dependence $S_p(\ell_m) \propto \ell_m^{\zeta_p}$ in the inertial interval. The gray region corresponds to the intermittent dissipation range (\ref{eq_DR4b}). Red circles mark the $p$-dependent cutoff scales (\ref{eq_DR4}). (b) Anomalous exponents $\zeta_p$ computed in the interval $-10 \le p \le 18$ with the step $\Delta p = 0.2$. The shaded area indicates error bounds, and the blue line corresponds to the K41 linear dependence $p/3$. Dashed lines indicate the slopes $h_{\max}$ and $h_{\min}$ at large negative and positive $p$. The results use simulations with $\mathrm{R} = 10^{10}$ and $T = 5\times 10^4$.}
\label{fig2}
\end{figure}

Let us introduce the structure function as
    \begin{equation}
    S_p(\ell_m) = \left\langle \mathcal{A}_m^p[u] \right\rangle_t,
    \quad p \in \mathbb{R},
    \label{eq1_Sp}
    \end{equation}
where $\langle \cdot \rangle_t$ denotes an average with respect to time $t \ge 0$. Numerical tests suggest that these averages do not depend on (generic) initial conditions at $t = 0$. Furthermore, the structure functions demonstrate accurate power-law scalings 
    \begin{equation}
    S_p(\ell_m) \propto \ell_m^{\zeta_p}
    \label{eq1_Sp2}
    \end{equation}
in the inertial interval for both positive and negative orders $p$ as shown in Fig.~\ref{fig2}(a). We remark that using the proper definition (\ref{eq_IRdef}) of the inertial interval increases the accuracy of measured exponents $\zeta_p$, as we explain later in Section~\ref{subsec_IDR}. The nonlinear dependence of exponents on $p$ shown in Fig.~\ref{fig2}(b) is a distinctive feature of  intermittency~\cite{frisch1999turbulence,l1998improved,biferale2003shell}. This property implies that all original scaling symmetries (\ref{eq_S1ideal}) are broken in the stationary statistics. In particular, intermittency breaks the K41 prediction $\zeta_p^{\mathrm{K41}} = p/3$. For this reason, exponents $\zeta_p$ are called anomalous. 

\subsection{Derivation of anomalous exponents from the hidden symmetry}\label{subsec_anom}

Following the derivations presented earlier in \cite{mailybaev2020hidden,mailybaev2022shell}, we now demonstrate that the anomalous power-law scaling of structure functions (\ref{eq1_Sp2}) follows from the hidden scale invariance.
As the first step, we express structure functions in terms of multipliers. Recalling our definition of multipliers (\ref{eqKm}), we express the velocity amplitude as the product 
    \begin{equation}
    \mathcal{A}_m[u] = u_0 \prod_{j = 0}^{m-1}\mathcal{X}_{-j}[U^{(m)}],
    \label{eq1_Sp2b}
    \end{equation}
where we used $\mathcal{A}_0[u] = u_0$ following from (\ref{eq2A}) and (\ref{eq2b}). Changing from the original time average $\langle \cdot \rangle_t$ to the average $\langle \cdot \rangle_{\tau^{(m)}}$ with respect to rescaled time $\tau^{(m)} \ge 0$, one can express the structure function (\ref{eq1_Sp}) as (see Appendix \ref{secA_3} for derivation)
    \begin{equation}
    S_p(\ell_m) 
    = u_0^p \, \frac{\Big\langle \Big(\prod_{j = 0}^{m-1}\mathcal{X}_{-j}[U^{(m)}]\Big)^{p-1} \Big\rangle_{\tau^{(m)}}
    }{\Big\langle \Big(\prod_{j = 0}^{m-1}\mathcal{X}_{-j}[U^{(m)}]\Big)^{-1} \Big\rangle_{\tau^{(m)}}}.
    \label{eq1_SpF}
    \end{equation}

Next we express time averages as integrals with respect to corresponding probability measures. For this purpose, we use the subsripts $\ominus$ and $-$ to denote the sequences
    \begin{equation}
	\mathbf{x}_\ominus = (x_{0},x_{-1},x_{-2},\ldots),\quad    
	\mathbf{x}_- = (x_{-1},x_{-2},\ldots).
	\label{eq1_SpMx}
    \end{equation}
Let $d\mu^{(m)}(\mathbf{x}_\ominus)$ be a probability measure describing the statistics of multipliers $\mathbf{x}_\ominus = \boldsymbol{\mathcal{X}}_\ominus[U^{(m)}]$ as functions of $\tau^{(m)}$. 
Expressing the product 
    \begin{equation}
	\prod_{j = 0}^{m-1}\mathcal{X}_{-j}[U^{(m)}] = \prod_{j = 0}^{m-1}x_{-j} 
	\label{eq1_SpMxB}
    \end{equation}
and using the ergodicity assumption, one writes Eq.~(\ref{eq1_SpF}) in the form
    \begin{equation}
    S_p(\ell_m) 
    = u_0^p\, \frac{\int \big(\prod_{j = 0}^{m-1}x_{-j}\big)^{p-1} d\mu^{(m)}(\mathbf{x}_\ominus)}{\int \big(\prod_{j = 0}^{m-1}x_{-j}\big)^{-1} d\mu^{(m)}(\mathbf{x}_\ominus)}. 
    \label{eq1_SpMd}
    \end{equation}
For further analysis it is convenient to write this expression as
    \begin{equation}
    S_p(\ell_m) 
    = u_0^p \int d\mu_p^{(m)}, 
    \label{eq1_SpM}
    \end{equation}
where $d\mu_p^{(m)}$ is a positive (generally not a probability) measure defined as
    \begin{equation}
    	    d\mu_p^{(m)}(\mathbf{x}_\ominus)
	    = \frac{1}{c_m} \Big(\prod_{j = 0}^{m-1}x_{-j}\Big)^{p-1}\, d\mu^{(m)}(\mathbf{x}_\ominus), \quad
	    c_m = \int
	    \Big(\prod_{j = 0}^{m-1}x_{-j}\Big)^{-1} d\mu^{(m)}(\mathbf{x}_\ominus).
    \label{eq1_SpMM}
    \end{equation}

Our goal now is to relate the measures $d\mu_p^{(m)}$ for different $m$. Let $p^{(m)}(x_1|\mathbf{x}_\ominus)$ be a conditional probability density of $x_1 = \mathcal{X}_1^{(m)}[U^{(m)}]$ given the values of multipliers $\mathbf{x}_\ominus = \boldsymbol{\mathcal{X}}_\ominus[U^{(m)}]$. Then, one can expresses the probability measure $d\mu_p^{(m+1)}(\mathbf{x}_\ominus)$ for the multipliers $\mathbf{x}_\ominus = \boldsymbol{\mathcal{X}}_\ominus[U^{(m+1)}]$ as (see Appendix \ref{secA_3} for the derivation)
    \begin{equation}
	    d\mu_p^{(m+1)}(\mathbf{x}_\ominus) 
	    = x_0^p \,p^{(m)} (x_0|\mathbf{x}_-)
	    \,dx_0 \,d\mu_p^{(m)}(\mathbf{x}_-).
    \label{eq1_SpM4}
    \end{equation}
Here $\mu_p^{(m)}(\mathbf{x}_-)$ denotes the image of measure $\mu_p^{(m)}(\mathbf{x}_\ominus)$ by the change (shift) of variables $\mathbf{x}_\ominus = (x_{0},x_{-1},\ldots) \mapsto \mathbf{x}_- = (x_{-1},x_{-2},\ldots)$.
It is convenient to introduce a linear operator $\mathcal{L}_p^{(m)}[d\mu]$ acting in the space of measures $d\mu(\mathbf{x}_\ominus)$ as
    \begin{equation}
    d\mu' = \mathcal{L}_p^{(m)}[d\mu],\quad 
	    d\mu'(\mathbf{x}_\ominus) 
	    = x_0^p \,p^{(m)} (x_0|\mathbf{x}_-)
	    \,dx_0 \, d\mu(\mathbf{x}_-).
    \label{eq1_SpM4m}
    \end{equation}
Then, Eq.~(\ref{eq1_SpM4}) takes the compact form
    \begin{equation}
	    d\mu_p^{(m+1)} = \mathcal{L}_p^{(m)}[d\mu_p^{(m)}].
    \label{eq1_SpM3}
    \end{equation}
Iterating this formula yields
    \begin{equation}
	    d\mu_p^{(m)} 
	    = \mathcal{L}_p^{(m-1)} \circ \mathcal{L}_p^{(m-2)} \circ \cdots \circ \mathcal{L}_p^{(1)}[d\mu_p^{(1)}].
    \label{eq1_SpM5}
    \end{equation}

So far, the derivations were general. Now we use the property of hidden scale invariance. It implies that the statistics of multipliers in the inertial range does not depend on the reference shell $m$. In particular, this independence refers to the conditional probability density $p^{(m)}$ and, hence, to the linear operator $\mathcal{L}_p^{(m)}$ as 
	\begin{equation}
	p^{(m)}(x_1|\mathbf{x}_\ominus) \approx \rho(x_1|\mathbf{x}_\ominus),
	\quad \mathcal{L}_p^{(m)} \approx \Lambda_p.
	\label{eq1_SpH9a}
	\end{equation}
Here $\rho(x_1|\mathbf{x}_\ominus)$ is the hidden-symmetric conditional density and $\Lambda_p$ is the corresponding operator expressed by Eq.~(\ref{eq1_SpM4m}) as
    \begin{equation}
    d\mu' = \Lambda_p[d\mu],\quad 
	    d\mu'(\mathbf{x}_\ominus) 
	    = x_0^p \,\rho (x_0|\mathbf{x}_-)
	    \,dx_0 \, d\mu(\mathbf{x}_-).
    \label{eq1_SpM4inf}
    \end{equation}
According to numerically studies of~\cite{benzi1993intermittency,eyink2003gibbsian,mailybaev2021hidden}, multipliers at distant shells become statistically independent (correlations decay exponentially at distant shell numbers). Hence, both the density  $\rho(x_1|\mathbf{x}_\ominus)$ and the operator $\Lambda_p$ can be approximated using a truncation of the sequence $\mathbf{x}_\ominus$ to a finite number of adjacent multipliers~\cite{mailybaev2022shell}.

For the final step, we notice that the linear operator $\Lambda_p$ is positive (mapping positive measures to positive measures). Hence, its spectral radius is given by a real positive (Perron--Frobenius) eigenvalue $\lambda_p$ satisfying the eigenvalue problem~\cite{lax2007linear,deimling2010nonlinear,mailybaev2020hidden}
	\begin{equation}
	\Lambda_p [d\nu_p] = \lambda_p \,d\nu_p.
	\label{eq1_SpH9}
	\end{equation}
The eigenvector $d\nu_p(\mathbf{x}_\ominus)$ is a positive measure defined up to a positive factor, which we normalize by the condition $\int d\nu_p = 1$. Under the non-degeneracy assumption (referring to strict positivity and compactness~\cite[Sec. 19.5]{deimling2010nonlinear}), the Perron-Frobenius eigenvalue $\lambda_p$ is larger than absolute values of all remaining eigenvalues. Hence, measures (\ref{eq1_SpM5}) in the inertial interval (for large $m$) have the asymptotic form 
	\begin{equation}
	    d\mu_p^{(m)} \approx C_p \lambda_p^m \, d\nu_p,
    	\label{eq1_SpH10}
    	\end{equation}
where the coefficient $C_p$ does not depend on $m$.
Substituting expression (\ref{eq1_SpH10}) into (\ref{eq1_SpM}) with $\ell_m = 2^{-m}\ell_0$, we recover the asymptotic power law for the structure function as
	\begin{equation}
	S_p(\ell_m) \approx C_p u_0^p \left(\frac{\ell_m}{\ell_0}\right)^{\zeta_p},\quad 
	\zeta_p = -\log_2 \lambda_p.
    	\label{eq1_SpH11}
    	\end{equation}

\begin{figure}[t]
\centering
\includegraphics[width=1\textwidth]{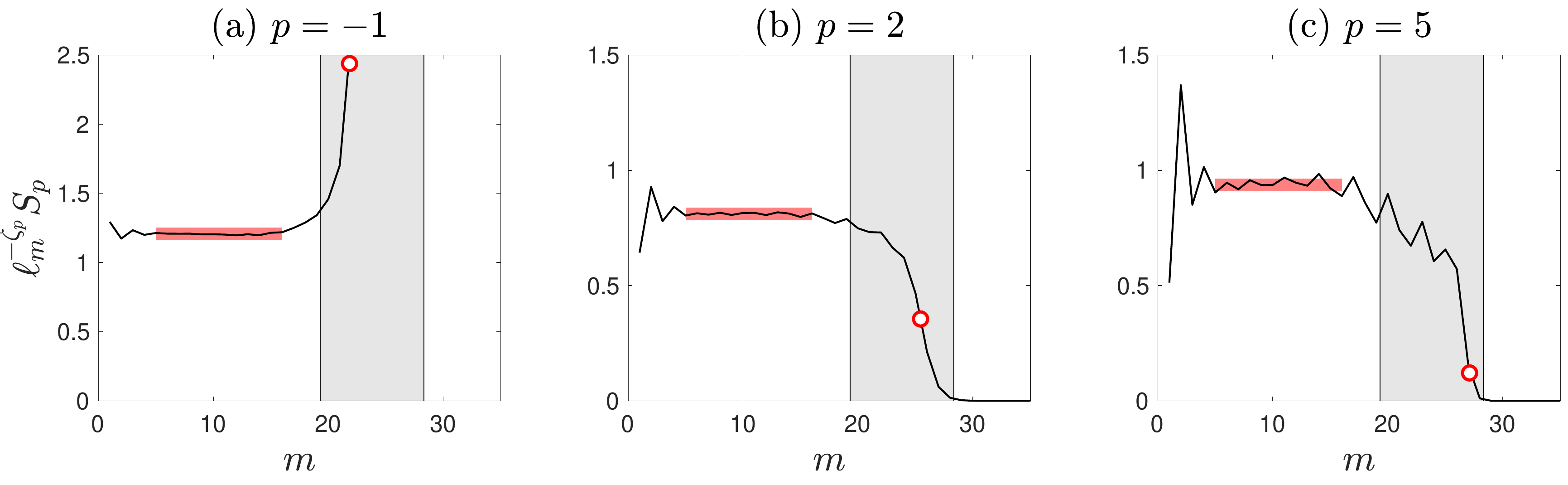}
\caption{Compensated structure functions $\ell_m^{-\zeta_p} S_p$ plotted as functions of the shell number $m$ for $p = -1,2,5$. These graphs confirm that the pre-factors $C_p$ in Eq.~(\ref{eq1_SpH11}) are constant in the inertial interval (indicated by bold red lines), but they change in the intermittent dissipation range (gray region). Red circles mark the $p$-dependent cutoff scales (\ref{eq_DR4}). The results use simulations with $\mathrm{R} = 10^{10}$ and $T = 5 \times 10^4$.}
\label{fig3}
\end{figure}

We derived the scaling exponents $\zeta_p$ in terms of the Perron-Frobenius eigenvalues $\lambda_p$. Generally, this relation yields the exponents depending nonlinearly on the order $p$, i.e., the intermittency~\cite{mailybaev2020hidden,mailybaev2021solvable}. Also, our derivation shows that the scaling laws are asymptotically precise, i.e., the pre-factors $C_p$ are constants independent of $m$. This property is verified numerically in Fig.~\ref{fig3}. According to Eq.~(\ref{eq1_SpM5}), pre-factors $C_p$ are determined by the statistics at small $m$. Hence, they depend on the forcing conditions. 

\begin{figure}[t]
\centering
\includegraphics[width=0.99\textwidth]{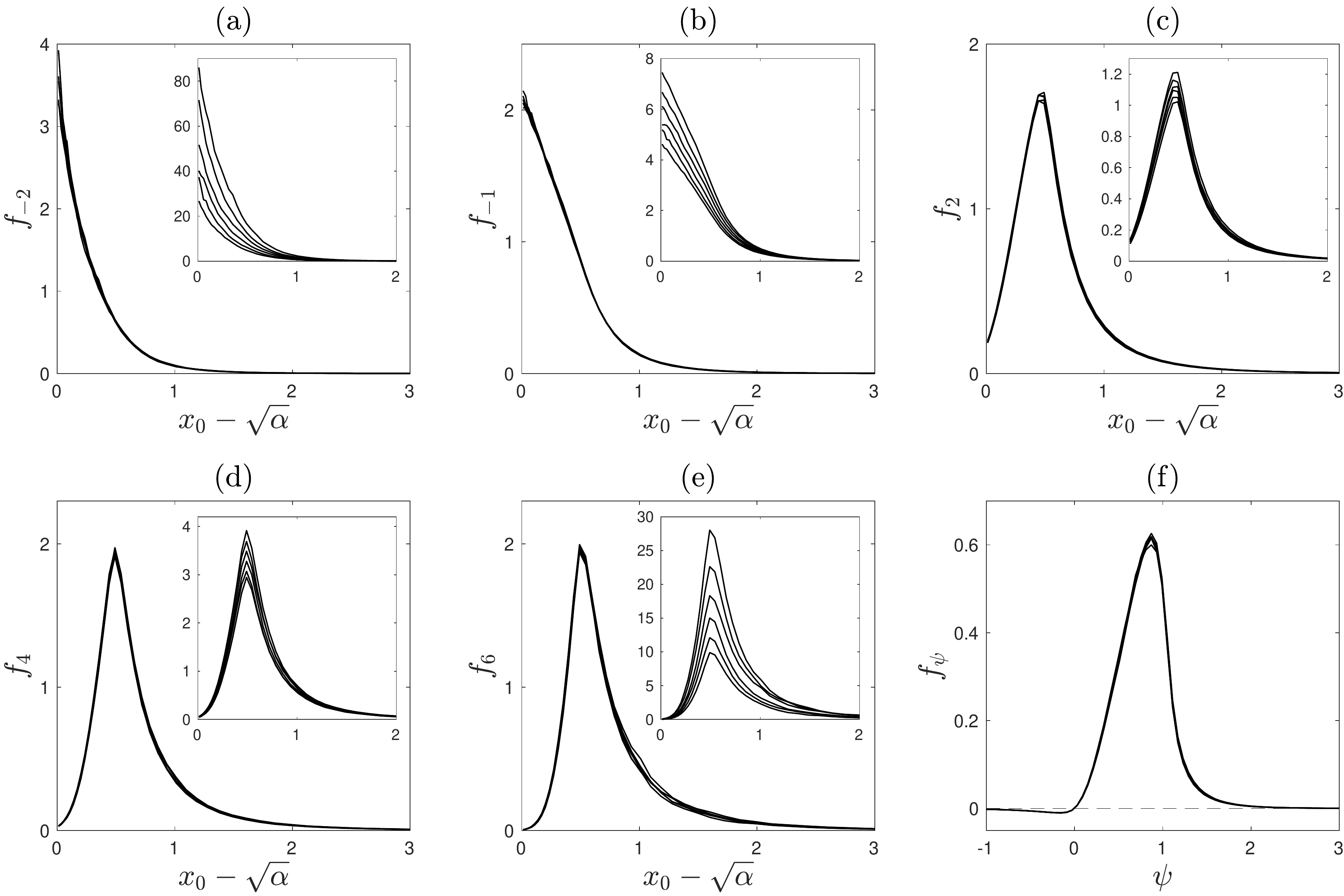}
\caption{(a-e) Marginal densities $f_p(x_0)$ from Eq.~(\ref{eq1_SpH10invM}) for $p = -2,-1,2,4$ and $6$. Each panel shows the collapse of six densities for reference shells $m = 8,\ldots,13$. 
Insets show the divergence of the same densities if anomalous exponents $\zeta_p$ in Eq.~(\ref{eq1_SpH10inv}) are replaced by the K41 values $p/3$. 
(f) Collapse of marginal densities $f_\psi(\psi)$ given by Eq.~(\ref{eq_EF1x}) for $m = 8,\ldots,13$.
The results use simulations with $\mathrm{R} = 10^{10}$ and $T = 5 \times 10^4$.}
\label{fig4}
\end{figure}

We have already confirmed numerically the hidden scale invariance for multipliers, which determine the universal linear operator (\ref{eq1_SpM4inf}).
It remains to verify relations (\ref{eq1_SpH10}) and (\ref{eq1_SpH11}) given by the Perron-Frobenius eigenmode (\ref{eq1_SpH9}). For this purpose, recalling that $\mathbf{x}_\ominus = (x_0,\mathbf{x}_-)$, we compute the marginal densities 
	\begin{equation}
	f_p(x_0) = \int d\nu_p(\mathbf{x}_\ominus) \, d \mathbf{x}_-,
    	\label{eq1_SpH10invM}
    	\end{equation}
where the measure $d\nu_p$ is approximated from Eqs.~(\ref{eq1_SpH10})--(\ref{eq1_SpH11}) as
	\begin{equation}
	d\nu_p \approx \frac{1}{C_p}\left(\frac{\ell_m}{\ell_0}\right)^{-\zeta_p}d\mu_p^{(m)}.
    	\label{eq1_SpH10inv}
    	\end{equation}
Equations (\ref{eq1_SpH9})--(\ref{eq1_SpH11}) are verified by showing that the densities $f_p(x_0)$ are independent of $m$ in the inertial interval for any fixed order $p$. 
The densities $f_p(x_0)$ are computed numerically as explained in Appendix~\ref{subsecA_num} and the results are presented in Fig.~\ref{fig4}(a-e) for $p = -2,-1,2,4,6$. Each panel shows six graphs for the shells $m = 8,\ldots,13$ in the middle of inertial interval.
The accurate collapse of these graphs provides the strong numerical support to our theory; see also \cite{mailybaev2022shell} for other numerical tests. In order to emphasize the high quality of the collapse, the insets of the same panels show how the densities diverge if $\zeta_p$ in Eq.~(\ref{eq1_SpH10inv}) are replaced by the K41 exponents $p/3$. 

\subsection{Scaling of generalized structure functions}
\label{subsec_genSF}

Here we generalize the previous results to a larger class of observables. Consider a sequence of real-valued (non necessarily positive) functions $\Psi_m[u]$ for integer numbers $m \in \mathbb{Z}$. We assume that these functions have the property of positive homogeneity of a given degree $p$:
	\begin{equation}
	\Psi_m[au] = a^p\Psi_{m}[u], \quad a > 0,
    	\label{eq_GS1}
    	\end{equation}
and satisfy the condition of scale invariance: 
	\begin{equation}
	\Psi_m[u] = \Psi_{m-1}[u'], \quad 
	u' = \left(u'_n\right)_{n\in\mathbb{Z}} = \left(u_{n+1}\right)_{n\in\mathbb{Z}}.
    	\label{eq_GS1Prime}
    	\end{equation}
The generalized structure function of order $p$ is defined as the time-averaged value 
	\begin{equation}
	S_\psi(\ell_m) = \langle \Psi_m[u] \rangle_t.
    	\label{eq_GS1b}
    	\end{equation}
Examples include both the standard structure functions $\langle |u_m|^p \rangle_t$ for $\Psi_m[u] = |u_m|^p$ and the structure functions (\ref{eq1_Sp}) for $\Psi_m[u] = \mathcal{A}_m^p[u]$. Another important example considered in the next Section~\ref{subsec_AD} is related to a flux of energy.
	
We now derive the asymptotic power law for the generalized structure function using the results of the previous section.
The transformation of Eq.~(\ref{eq_GS1b}) to rescaled variables yields the expression (see Appendix \ref{secA_3})
	\begin{equation}
	S_\psi(\ell_m) = u_0^p \int \psi p_\psi^{(m)}(\psi|\mathbf{x}_\ominus)\, d\psi\, d\mu_p^{(m)}(\mathbf{x}_\ominus),
    	\label{eq_GS2}
    	\end{equation}
where the measure $d\mu_p^{(m)}(\mathbf{x}_\ominus)$ is defined in (\ref{eq1_SpMM}), and $p_\psi^{(m)}(\psi|\mathbf{x}_\ominus)$ denotes a conditional probability density for the variable $\psi = \Psi_0[U^{(m)}]$ given the values of multipliers $\mathbf{x}_\ominus = \boldsymbol{\mathcal{X}}_\ominus[U^{(m)}]$. The hidden scale invariance implies that the densities
	\begin{equation}
	p_\psi^{(m)}(\psi|\mathbf{x}_\ominus) \approx \rho_\psi(\psi|\mathbf{x}_\ominus) 
    	\label{eq_GS4D}
    	\end{equation}
do not depend on the reference shell $m$ within the inertial interval. Combining this property with Eqs.~(\ref{eq1_SpH10}), (\ref{eq1_SpH11}) and (\ref{eq_GS2}), yields
	\begin{equation}
	S_\psi(\ell_m) \approx I_\psi C_p u_0^p \left(\frac{\ell_m}{\ell_0}\right)^{\zeta_p},\quad
	I_\psi = \int \psi \rho_\psi(\psi|\mathbf{x}_\ominus)\, d\psi\, d\nu_p(\mathbf{x}_\ominus),
    	\label{eq_GS4}
    	\end{equation}
provided that the integral $I_\psi$ is finite and nonzero.

We remark that Eq.~(\ref{eq_GS4}) yields not only the scaling law but also a pre-factor, where both the exponent $\zeta_p$ and the forcing-dependent coefficient $C_p$ are the same for all generalized structure functions of a given order $p$. We would like to stress the importance of the nondegeneracy condition, $0 < |I_\psi| < \infty$.
For example, in the standard definition of structure functions, one takes $\Psi_m[u] = |u_m|^p$ and Eq.~(\ref{eq_GS4}) yields $\langle|u_m|^p\rangle_t \propto  \ell_m^{\zeta_p}$. This relation, however, does not hold for  $p \le -2$, since the corresponding integral $I_\psi$ and structure function diverge. If $I_\psi = 0$, then the power law in Eq.~(\ref{eq_GS4}) vanishes. However, the asymptotic power-law scaling of $S_\psi(\ell_m)$ may still exist: it can be related to the next (after the Perron-Frobenius) leading mode of the linear operator $\Lambda_p$; see Eqs.~(\ref{eq1_SpM5}) and (\ref{eq1_SpH9a}). This argument also applies if $C_p = 0$, which may occur for a special form of forcing. 

As a final remark, let us mention integrated multi-time correlation functions considered in \cite{l1997temporal,biferale1999multi,mitra2004varieties}. We expect that our generalized characterization of scaling laws can further be extended to such observables.

\subsection{Hidden symmetry of anomalous dissipation}
\label{subsec_AD}

As another example of generalized structure function, let us consider a flux of energy.
Multiplying both sides of the inertial-interval equation (\ref{eq1Euler}) by $u_n^*$ and taking real part, after some elementary manipulations using (\ref{eq1Bn}), one derives the local energy balance
	\begin{equation}
	\frac{d}{dt}\frac{|u_n|^2}{2} = \Pi_n-\Pi_{n+1}.
    	\label{eq_EF2}
    	\end{equation}
Here $\Pi_n$ is the energy flux from shell $n-1$ to $n$ given by
	\begin{equation}
	\Pi_n[u] = k_n\,\mathrm{Im}\left(
	u_{n+1}u_n^*u_{n-1}^*
	+\frac{u_nu_{n-1}^*u_{n-2}^*}{4} \right).
    	\label{eq_EF1}
    	\end{equation}
One can see that $\Psi_m[u] = \Pi_m[u]/k_m$ is a generalized structure function of order $p = 3$ satisfying conditions (\ref{eq_GS1}) and (\ref{eq_GS1Prime}). 

Using Eq.~(\ref{eq_GS4}) with $p = 3$ we obtain 
	\begin{equation}
	\langle \Pi_m[u] \rangle_t 
	=  k_m \langle \Psi_m[u] \rangle_t \approx I_\psi C_3 \frac{u_0^3}{\ell_0} \left(\frac{\ell_m}{\ell_0}\right)^{\zeta_3-1},
    	\label{eq_EF1xx}
    	\end{equation}
where we expressed $k_m = 1/\ell_m$. If $I_\psi C_3 \ne 0$, then $\zeta_3 = 1$ is the only exponent compatible with the condition $\langle \Pi_m[u]\rangle_t  = \langle \Pi_{m+1}[u] \rangle_t$ following from Eq.~(\ref{eq_EF2}). This fact, well-known by the name of \textit{dissipation anomaly} \cite{eyink2006onsager}, implies that the average flux of energy approaches an asymptotic value 
	\begin{equation}
	\langle \Pi_m[u] \rangle_t \approx \frac{I_\psi C_3u_0^3}{\ell_0}
    	\label{eq_EF1_DA}
    	\end{equation}
in the inertial interval, which is independent of Reynolds number. 

We verify the hidden symmetry property (\ref{eq_GS4D}) numerically by considering the functions 
	\begin{equation}
	f_\psi(\psi) = \int \psi \rho_\psi(\psi|\mathbf{x}_\ominus)\,d\nu_3 (\mathbf{x}_\ominus),
    	\label{eq_EF1x}
    	\end{equation}
where $\rho_\psi$ and $d\nu_3$ are approximated, respectively, using $p_\psi^{(m)}$ and Eq.~(\ref{eq1_SpH10inv}) with $\zeta_3 = 1$; see Appendix~\ref{subsecA_num} for more details. 
Figure \ref{fig4}(f) shows the functions $f_\psi(\psi)$ computed for different $m$. The accurate collapse of these functions verifies our conclusions based on the hidden symmetry and yields $I_\psi = \int f_\psi\, d\psi \approx 0.41$. The latter provides a positive value to the energy flux (\ref{eq_EF1_DA}).


\subsection{Large deviation theory and multifractality} \label{subsec_ldt}

Our derivation of anomalous power laws for structure functions has much in common with the large deviation theory for a Markov process~\cite{benzi1993intermittency,frisch1999turbulence,eyink2003gibbsian}. The analogy becomes transparent if one takes a logarithm of the multiplicative relation (\ref{eq1_Sp2b}) divided by $m$. After elementary manipulations, this yields
    \begin{equation}
    W_m = -\frac{1}{m}\log_2 \frac{\mathcal{A}_m[u]}{u_0} = \frac{w_1+w_2+\cdots+w_m}{m},
    \label{eq_LD1cc}
    \end{equation}
where 
    \begin{equation}
    w_n = -\log_2\mathcal{X}_{n-m}[U^{(m)}] = -\log_2 \frac{\mathcal{A}_n[u]}{\mathcal{A}_{n-1}[u]}
    \label{eq_LD1ccM}
    \end{equation}
is a negative logarithm of the multiplier (\ref{eqKm}).
Considering $w_1,\ldots,w_{m}$ as random variables, one identifies $W_n$ with their sample mean. The hidden symmetry assumption (\ref{eq1_SpH9a}) implies that the probability of the next variable $w_{m+1}$ conditioned on $w_{m},w_{m-1},\ldots$ does not depend on $m$. Also, these variables become statistically independent at distant shells~\cite{benzi1993intermittency,eyink2003gibbsian,mailybaev2022shell}. Hence, the  sequence $w_1,w_2,\ldots$ has properties of a generalized Markov chain. 
Notice, however, that the hidden scale invariance has an intrinsic feature distinguishing it from a Markov chain: the change $m \mapsto m+1$ must also be accompanied by the change of rescaled time, $\tau^{(m)} \mapsto \tau^{(m+1)}$. 

We now derive the Parisi--Frisch multifractal (phenomenological) theory of turbulence~\cite{frisch1985singularity,frisch1999turbulence} as a large deviation theory following from the hidden scale invariance. Indeed, combining Eq.~(\ref{eq_LD1cc}) and the power-law scaling (\ref{eq1_SpH11}) of structure functions (\ref{eq1_Sp}), one obtains
    \begin{equation}
    \frac{\left\langle \mathcal{A}_m^p[u] \right\rangle_t}{u_0^p} 
    = \left\langle 2^{-mpW_m} \right\rangle_t \approx 2^{-m\zeta_p} C_p.
    \label{eq_LD1cX}
    \end{equation}
For convenience, we use here exponential functions with base $2$ and recall that $\ell_m/\ell_0 = 2^{-m}$. 
The crucial property of velocity amplitudes $\mathcal{A}_m[u]$ is that they define exponents $\zeta_p$ for all $p \in \mathbb{R}$; see Fig.~\ref{fig2}.
In this case, one can apply the G\"artner--Ellis Theorem; see \cite[\S3.3.1]{touchette2009large} and Appendix~\ref{subsec_LDT}. This theorem yields the Large Deviation Principle for large $m$ formulated as the probability 
    \begin{equation}
    P\left(W_m \in [h,h+dh] \right) \approx 2^{-mJ(h)}\, dh.
    \label{eq_LD1}
    \end{equation}
Here $J(h)$ is the rate (Cram\'er) function defined as
    \begin{equation}
    J(h) = \sup_{p \in \mathbb{R}} \, \left(\zeta_p-ph\right).
    \label{eq_LD1bJ}
    \end{equation}
Since $\zeta_p$ is a concave function of $p$, relation (\ref{eq_LD1bJ}) is solved implicitly as
    \begin{equation}
    h = \frac{d \zeta_p}{dp}, \quad J = \zeta_p-p\,\frac{d \zeta_p}{dp}.
    \label{eq_LD1bJimp}
    \end{equation}
The inverse of (\ref{eq_LD1bJ}) reads (see Appendix~\ref{subsec_LDT})
    \begin{equation}
    \zeta_p = \inf_{h \in \mathbb{R}} \, \big(ph+J(h)\big).
    \label{eq_LD1inv}
    \end{equation}

Using relations (\ref{eq_LD1cc}) and $\ell_m/\ell_0 = 2^{-m}$, we express
    \begin{equation}
    W_m = h \ \ \Rightarrow \ \ \mathcal{A}_m[u] = u_0\left(\frac{\ell_m}{\ell_0}\right)^{h}.
    \label{eq_LD1bX}
    \end{equation}
Then, the large deviation principle (\ref{eq_LD1}) is written (less formally) as
    \begin{equation}
    \mathcal{A}_m[u] \sim u_0\left(\frac{\ell_m}{\ell_0}\right)^{h}\ \ \textrm{with probability} \ \ 
    P \sim \left(\frac{\ell_m}{\ell_0}\right)^{J(h)}.
    \label{eq_LD1b}
    \end{equation}

The multifractal theory follows, if one identifies $J(h)$ with a (fractal) codimension of a subset corresponding to the scaling law $\mathcal{A}_m[u] \propto \ell_m^h$. Indeed, as in the multifractal model~\cite{frisch1985singularity,frisch1999turbulence}, one represents the averaged moment of $\mathcal{A}_m[u]$ in the form
    \begin{equation}
    \langle \mathcal{A}_m^p[u] \rangle_t \sim u_0^p \int \left(\frac{\ell_m}{\ell_0}\right)^{ph+J(h)}\, d\mu(h),
    \label{eq_LD1c}
    \end{equation}
where the exponents $ph$ and $J(h)$ are due to the velocity amplitude and probability in Eq.~(\ref{eq_LD1b}), and $d\mu(h)$ measures a contribution of different $h$. Then, the power law $\langle \mathcal{A}_m^p[u] \rangle_t \propto \ell_m^{\zeta_p}$ is given by the smallest exponent $ph+J(h)$ provided by Eq.~(\ref{eq_LD1inv}).

\begin{figure}[t]
\centering
\includegraphics[width=0.85\textwidth]{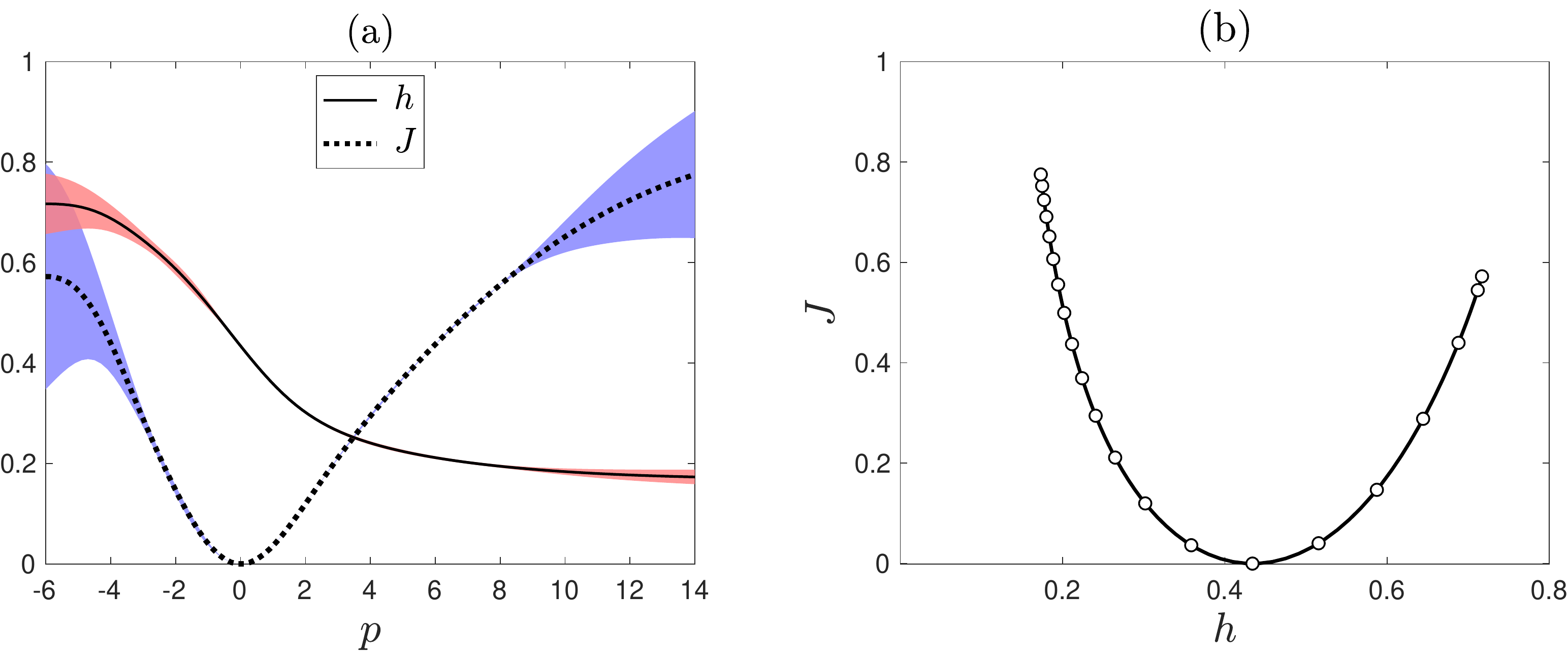}
\caption{(a) Scaling exponent $h$ and fractal codimension $J$ given by Eq.~(\ref{eq_LD1bJimp}) as functions of $p$. The shaded area indicates error bounds. (b) The resulting function $J(h)$. Circles correspond to integer values of $p = -6,\ldots,14$. The results use simulations with $\mathrm{R} = 10^{10}$ and $T = 5 \times 10^4$.}
\label{fig2h}
\end{figure}

Figure~\ref{fig2h}(a) shows graphs of expressions (\ref{eq_LD1bJimp}) computed numerically using the exponents from Fig.~\ref{fig2}(b), and Fig.~\ref{fig2h}(b) presents the resulting function $J(h)$.
Error estimates shown in panel (a) become large for large (negative and positive) orders $p$. We remark that these errors are not only due to statistical fluctuations, but also due to oscillations emerging from the forcing and dissipation regions; see Section~\ref{sec_visc}. Even though large errors hinder the analysis of large orders, Fig.~\ref{fig2}(b) suggests that the asymptotic dependence of $\zeta_p$ is linear for large $|p|$. This implies that the exponents $h$ corresponding to finite rates $J(h)$ have finite lower and upper limits. We estimate them numerically as
    \begin{equation}
	h_{\min} \le h \le h_{\max}, \quad 
	h_{\min} \approx 0.173 \pm 0.015, \quad 
	h_{\max} \approx 0.72 \pm 0.06. 
    \label{eq_Lminmax}
    \end{equation}
Using the exponent $h_{\max}$ in Eq.~(\ref{eq_LD1b}), one obtains the steepest decay of velocity amplitudes in the inertial interval as
    \begin{equation}
    |u_m| \sim \mathcal{A}_m[u] \sim u_0\left(\frac{\ell_m}{\ell_0}\right)^{h_{\max}} = u_0 2^{-mh_{\max}}.
    \label{eq_Lmax2}
    \end{equation}
It follows that the sums with respect to $j$, which appear in Eq.~(\ref{eq2A}) and other similar expressions, converge exponentially in the inertial interval for any $\alpha < 4^{-h_{\max}} \approx 0.37$. Recall that our numerical simulations use $\alpha = 1/8$.

\section{Breaking of the hidden symmetry by dissipation and forcing}
\label{sec_visc}

\begin{figure}[t]
\centering
\includegraphics[width=0.85\textwidth]{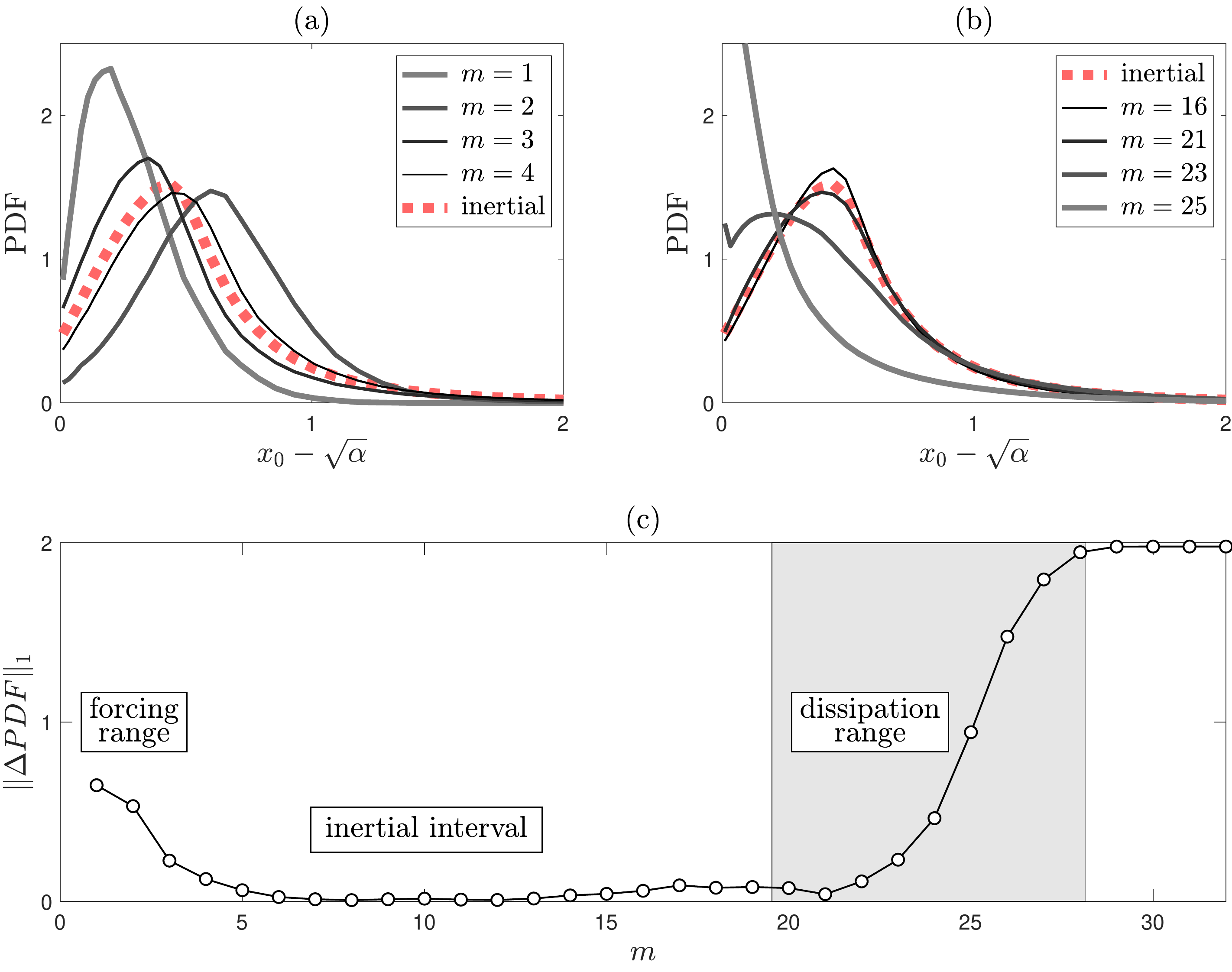}
\caption{Divergence of PDFs of multipliers $x_0 = \mathcal{X}_0[U^{(m)}]$ from the hidden-symmetric form (red dotted line) for shells in (a) forcing and (b) dissipation ranges. (c) $L^1$-norm of a difference between the multiplier PDF at shell $m$ and the hidden-symmetric PDF. The results use simulations with $\mathrm{R} = 10^{10}$ and $T = 10^4$.}
\label{fig5}
\end{figure}

In this section we investigate how the hidden scale invariance is broken by the dissipation at small scales and by the forcing at large scales. For this study, we consider PDFs of multipliers $x_0 = \mathcal{X}_0[U^{(m)}]$ as observables at different reference shells $m$. Figure \ref{fig5} presents these PDFs for large forcing scales (panel a) and small dissipation scales (panel b) compared to the hidden-symmetric PDF from the inertial interval. One can see the divergence of PDFs as the reference shell moves away from the inertial interval. For a global picture, we present in the panel (c) the $L^1$ norm (integrated absolute value) of a difference between the PDF at shell $m$ and the hidden-symmetric PDF from the inertial interval. One can see that this norm vanishes in the inertial interval featuring the hidden-symmetric statistics. Below in this section we analyze separately how the hidden symmetry is broken in its left (forcing) and right (dissipation) sides.

\subsection{Forcing range}
\label{subsec_FR}

\begin{figure}[t]
\centering
\includegraphics[width=1\textwidth]{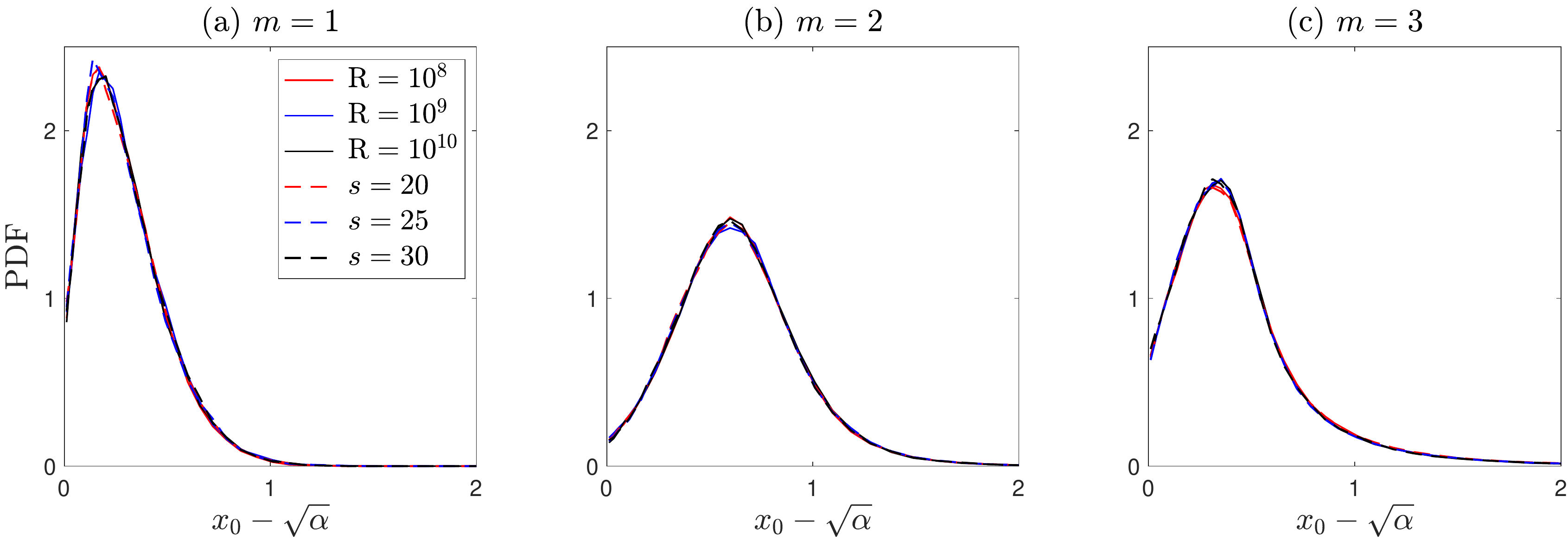}
\caption{Collapse of PDFs of the multipliers $x_0 = \mathcal{X}_0[U^{(m)}]$ for reference shells $m = 1,2,3$ in the forcing range. Six curves in each panel correspond to simulations with $\mathrm{R} = 10^8, 10^9, 10^{10}$ and $T = 10^4$, as well as to the viscous cutoff models with $s = 20,25,30$ considered in Section~\ref{sec_diss}. }
\label{fig5b}
\end{figure}

\begin{figure}[t]
\centering
\includegraphics[width=0.85\textwidth]{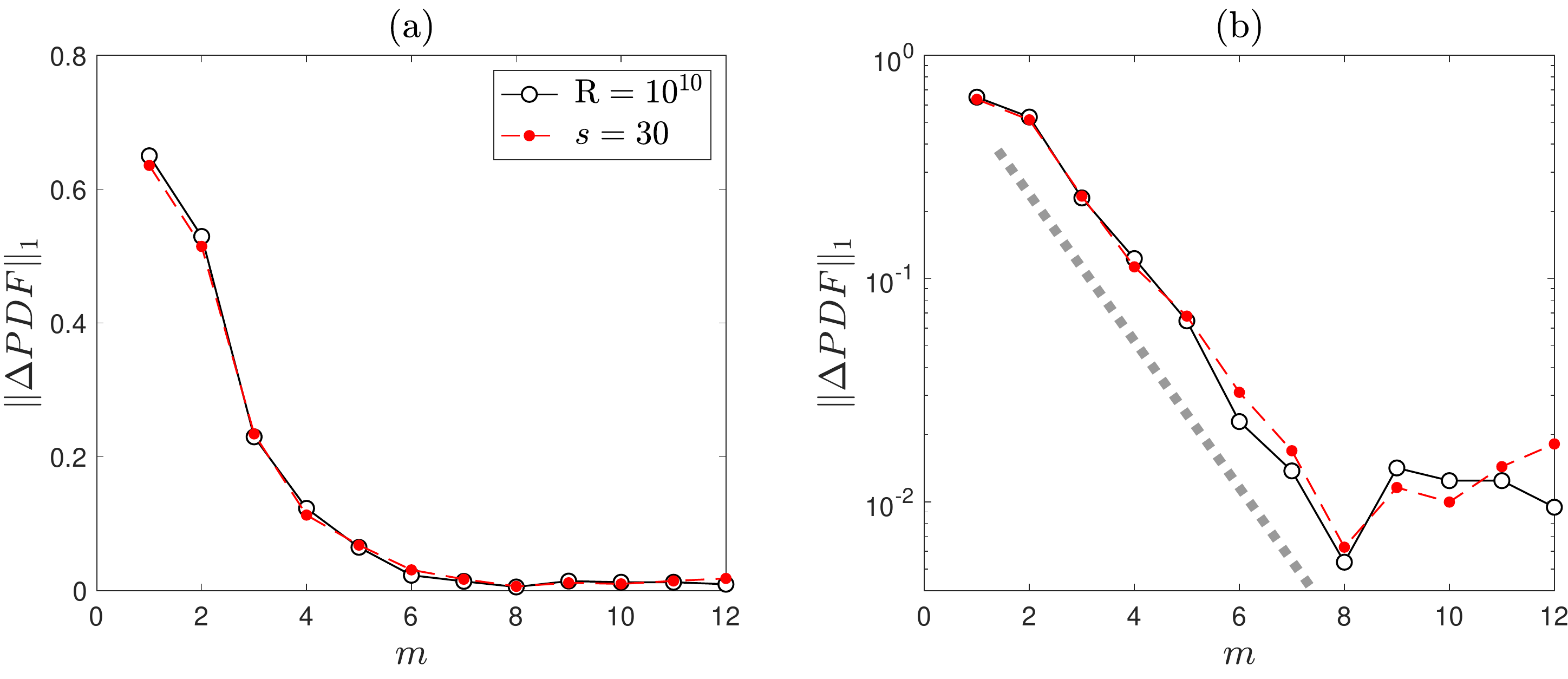}
\caption{$L^1$-norm of the difference between PDFs of the multipliers $x_0 = \mathcal{X}_0[U^{(m)}]$ at shells $m = 1,\ldots,12$ and the hidden-symmetric PDF: (a) linear and (b) logarithmic vertical scale. Solid lines correspond to $\mathrm{R} = 10^{10}$, and dashed lines to the viscous cutoff model with $s = 30$. The dotted gray line indicates an exponential dependence.}
\label{fig5c}
\end{figure}

It is known from numerical simulations that the large-scale statistics depends on forcing (given in our case by the boundary conditions with parameters $\ell_0$ and $u_0$), but does not depend on viscosity for very large Reynolds numbers. A similar conclusion follows for the rescaled formulation, in which the boundary conditions take the form (\ref{eq1R_bc}). Figure~\ref{fig5b} confirms that PDFs of multipliers converge as $\mathrm{R} \to \infty$. Figure~\ref{fig5c}(a) demonstrates another type of convergence: PDFs of multipliers approach the hidden-symmetric form with increasing reference shell $m$. It follows from Fig.~\ref{fig5c}(b) presenting the same graph in vertical logarithmic scale that the convergence is exponential in $m$ in the region $2 \le m \le 8$:
    \begin{equation}
    \label{eq_Lexp}
    \|\Delta\,PDF\|_1 \propto 2^{-\zeta_F m} \propto \ell_m^{\zeta_F}, \quad
    \zeta_F \approx 1.1.
    \end{equation}

Relation (\ref{eq_Lexp}) suggests a natural interpretation of the observed statistical properties in the forcing range. Namely, the exponential decay features a leading (slowest) mode with the Lyapunov exponent $-\zeta_F$ in a transition from the boundary state at scale $\ell_0$ to the stable hidden-symmetric state at small scales $\ell_m \ll \ell_0$. Theoretical understanding of this transition requires a consistent stability theory of the hidden-symmetric state, whose development would be an interesting direction for further research.

\begin{figure}[t]
\centering
\includegraphics[width=1\textwidth]{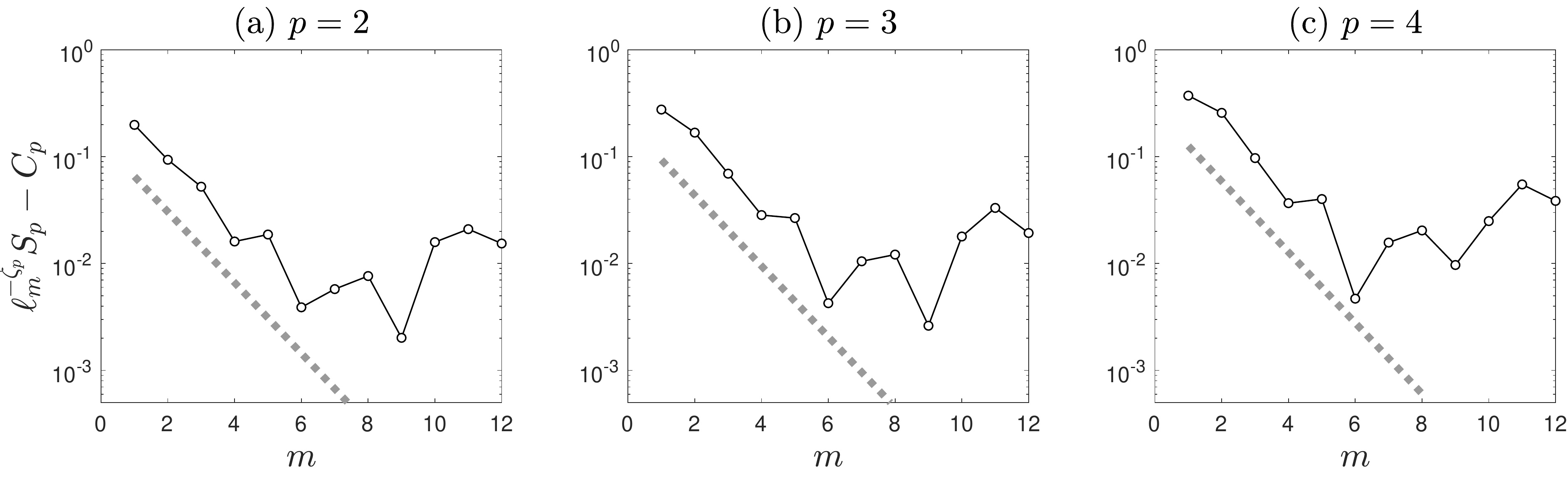}
\caption{Deviations of structure functions from the power law in the forcing range for $p = 2,3,4$ in logarithmic vertical scale. The dashed line shows the exponential mode $\propto 2^{-\zeta_F m}$. }
\label{fig3b}
\end{figure}

Finally, let us consider the statistics of original shell velocities characterized by structure functions $S_p(\ell_m) = \left\langle \mathcal{A}_m^p[u] \right\rangle_t$. We showed in Section~\ref{subsec_anom} that the power law (\ref{eq1_SpH11}) in the inertial interval is a consequence of the two limits: convergence of the multiplier statistics to the hidden-symmetric state (\ref{eq1_SpH9a}) and the subsequent convergence of the measures $d\mu_p^{(m)}$ to the Perron--Frobenius state (\ref{eq1_SpH10}). Therefore, the convergence of structure functions to power laws from the forcing side is a complicated process, whose better understanding requires the stability theory already mentioned in the previous paragraph. We visualize this convergence by plotting a discrepancy in the asymptotic relation (\ref{eq1_SpH11}) defined as $\ell_m^{-\zeta_p}S_p-C_p$, where we used $\ell_0 = u_0 = 1$. Such graphs for $p = 2,3,4$ are presented in Fig.~\ref{fig3b} together with the exponential mode (dotted line) from Fig.~\ref{fig5c}(b).  

\subsection{Intermittent dissipation range}
\label{subsec_IDR}

The self-similarity of the inertial interval is manifested in the rescaled formulation (\ref{eq1Ra}).
Let us now investigate the role of viscous dissipation in this system. 
It turns out that rescaled viscous terms are neither localized at specific shells, nor they have a closed form in terms of rescaled variables. The latter is because the intermittent Reynolds numbers (\ref{eq1Rb}) are proportional to amplitudes $\mathcal{A}_m[u]$. The consequence is a complicated structure of the dissipation range, which we now describe.

The rescaled formulation is designed by setting the reference shell $m$ at a scale of interest. Then the local statistics is described in terms of time $\tau^{(m)}$ and variables $U_N^{(m)}$, where the index $N$ takes zero or moderate values. The hidden self-similarity considered in Section~\ref{sec_intII} follows under the assumption that the viscous term in Eq.~(\ref{eq1Ra}) is negligible, i.e., that the local Reynolds number
    \begin{equation}
	\mathrm{R}_m[u] = \frac{\mathcal{A}_m[u]\ell_m}{\nu} \gg 1
	\label{eq_DR1L}
    \end{equation}
given by Eq.~(\ref{eq1Rb}) is large.
Using Eq.~(\ref{eq_LD1b}) of the Large Deviation Principle in Eq.~(\ref{eq_DR1L}), we have 
    \begin{equation}
	\mathrm{R}_m[u] \sim \frac{\ell_mu_0}{\nu}\left(\frac{\ell_m}{\ell_0}\right)^{h} = \mathrm{R}\,
	\left(\frac{\ell_m}{\ell_0}\right)^{1+h}
	\ \ \textrm{with probability} \ \ 
    	P \sim \left(\frac{\ell_m}{\ell_0}\right)^{J(h)},
	\label{eq_DR1h}
    \end{equation}
where $\mathrm{R} = \ell_0u_0/\nu$. Recall that the exponent $h$ varies in a finite interval (\ref{eq_Lminmax}). Hence, combining Eqs.~(\ref{eq_DR1L}) and (\ref{eq_DR1h}), we find that dissipation effects are negligible at all times for the scales 
    \begin{equation}
	\frac{\ell_m}{\ell_0} \gg \mathrm{R}^{-1/(1+h_{\max})}
	\label{eq_DR1max}
    \end{equation}
with $1/(1+h_{\max}) \approx 0.58$.
Similarly, dissipation effects are dominant at all times for very small scales 
    \begin{equation}
	\frac{\ell_m}{\ell_0} \ll \mathrm{R}^{-1/(1+h_{\max})}
	\label{eq_DR1min}
    \end{equation}
with $1/(1+h_{\min}) \approx 0.85$. 

\begin{figure}[t]
\centering
\includegraphics[width=0.8\textwidth]{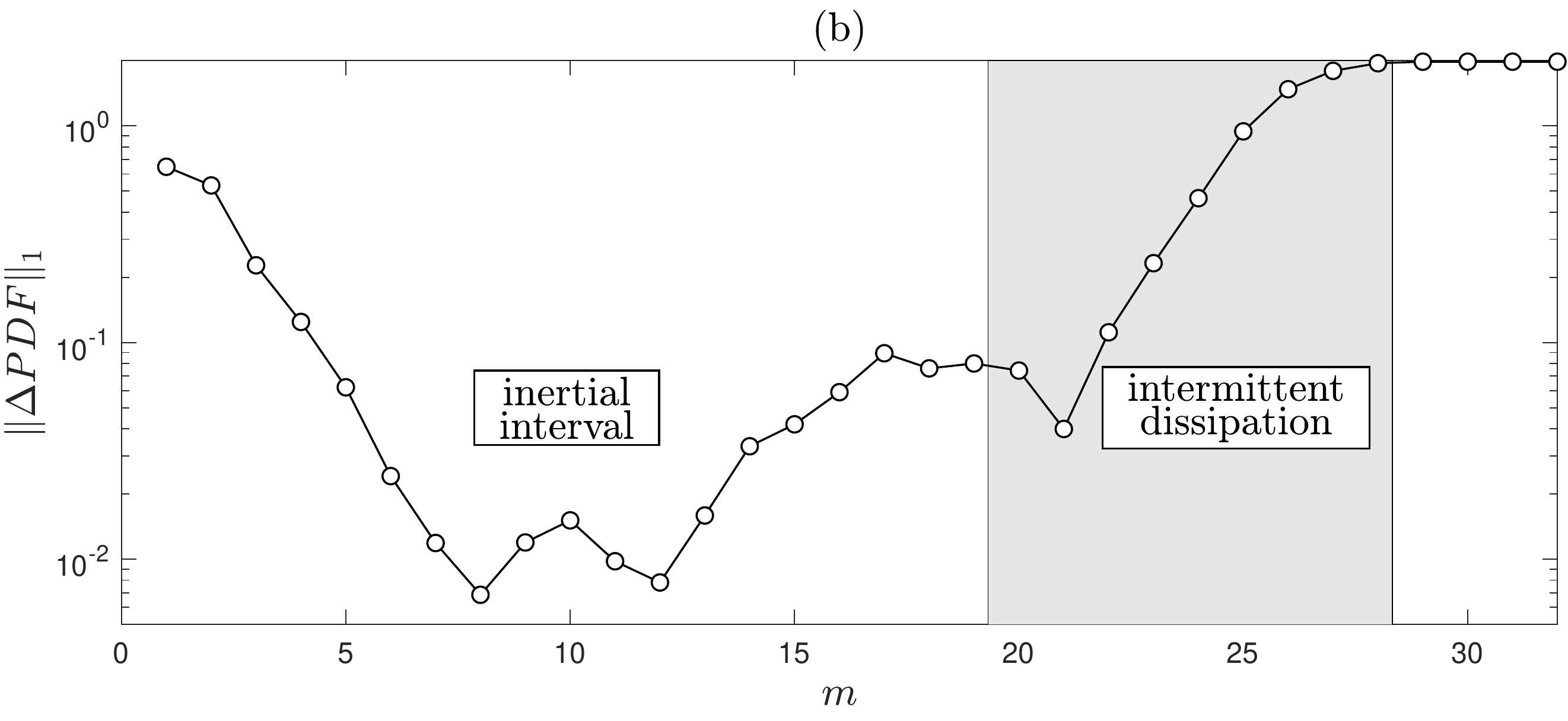}
\caption{Difference between the PDF of multiplier $x_0 = \mathcal{X}_0[U^{(m)}]$ at shell $m$ and the hidden-symmetric PDF presented in vertical logarithmic scale. The results use a simulation with $\mathrm{R} = 10^{10}$ and $T = 10^4$.}
\label{fig6}
\end{figure}

Let us consider the range  
    \begin{equation}
    \mathrm{R}^{-\frac{1}{1+h_{\min}}}
    \lesssim \frac{\ell_m}{\ell_0} \lesssim 
    \mathrm{R}^{-\frac{1}{1+h_{\max}}},
    \label{eq_DR4b}
    \end{equation}
which separates the larger scales (\ref{eq_DR1max}) with negligible dissipation and very small scales (\ref{eq_DR1max}) with dominant dissipation. At  scales of this range, local Reynolds numbers (\ref{eq_DR1h}) are either small or large with certain  probabilities, i.e., the viscous dissipation acts intermittently \cite{frisch1993prediction}. For this reason, we will refer to scales (\ref{eq_DR4b}) as the intermittent dissipation range. 
A consequence of this dissipative intermittency is the gradual breaking of the hidden symmetry within the whole range (\ref{eq_DR4b}). This is indeed confirmed in 
Fig.~\ref{fig5}(c) for the statistics of multipliers $x_0 = \mathcal{X}_0[U^{(m)}]$, which shows that a difference between the multiplier PDF and the hidden-symmetric PDF grows continuously from zero to a constant value in the gray region (\ref{eq_DR4b}). The constant attained at small scales (\ref{eq_DR1min}) is equal to $2$ because the multiplier distribution approaches the Dirac delta with $x_0 \approx 0$. Figure~\ref{fig6} shows the same graph but in vertical logarithmic scale. One can notice from both Figs.~\ref{fig5}(c) and \ref{fig6} that the deviations from the hidden-symmetric state are only moderately small at scales $\ell_m/\ell_0 \sim \mathrm{R}^{-1/(1+h_{\max})}$ ($m \approx 19$). The ultimate relaxation to the hidden-symmetric statistics requires a few extra shells,  e.g., $14 \le m \le 18$ in Fig.~\ref{fig6}. 

Our analysis suggests a natural definition of the inertial interval: it is a range of scales, at which the hidden scale invariance is restored in a statistical sense. At small scales, the hidden symmetry condition is given by Eq.~(\ref{eq_DR1max}). Thus, we identify the inertial interval with the scales (\ref{eq_IRdef}), as stated in the beginning of Section~\ref{sec_intII}.

\subsection{Structure functions in the intermittent dissipation range}
\label{subsec_IDRb}

Now let us study viscous effects in the statistics of original velocities, which we observe using the structure functions $S_p(\ell_m)$. Here our analysis resembles the phenomenological description of the intermediate dissipation range introduced in \cite{frisch1993prediction,bowman2006links}. We remark that though the concepts of ``intermittent'' and ``intermediate'' dissipation ranges are strongly related, the first refers to the breaking of hidden symmetry while the latter refers to (and depends on) specific order $p$ of the structure function.

Following \cite{frisch1993prediction}, we observe that a dominant contribution to the structure function (\ref{eq_LD1c}) of a given order $p$ comes from amplitude fluctuations (\ref{eq_LD1b}) of the form
    \begin{equation}
    \mathcal{A}_m[u] \sim u_0\left(\frac{\ell_m}{\ell_0}\right)^{H(p)}, \quad H(p) = \frac{d\zeta_p}{dp}.
    \label{eq_DR3}
    \end{equation}
Here $h = H(p)$ provides the minimum exponent $ph+J(h)$ as follows from Eqs.~(\ref{eq_LD1bJimp}) and (\ref{eq_LD1inv}) for a concave function $\zeta_p$. It follows from Eqs.~(\ref{eq_DR1L}) and (\ref{eq_DR1h}) that viscous terms are negligible for fluctuations with $h = H(p)$ if 
    \begin{equation}
    \ell_m \gg \eta(p) = \mathrm{R}^{-\frac{1}{1+H(p)}}\ell_0.
    \label{eq_DR4}
    \end{equation}
This condition defines a family of viscous cutoffs $\ell_m \sim \eta(p)$ depending on the order $p$ of a structure function. These cutoffs are clearly seen in Fig.~\ref{fig2}(a), where they correspond to the breakdown of power-law dependence $S_p(\ell_m) \propto \ell_m^{\zeta_p}$. At smaller scales $\ell_m \ll \eta(p)$ (larger shell numbers), shell velocities decay to zero at a much faster rate~\cite{frisch1993prediction}. Neglecting these velocities in the sum of Eq.~(\ref{eq2A}), one can show that our structure functions have a monofractal power-law scaling $S_p(\ell_m) = \left\langle \mathcal{A}_m^p[u] \right\rangle_t \propto \alpha^{pm/2}$ for $\ell_m \ll \eta(p)$; see Fig.~\ref{fig2}(a). 

The extension of power laws till the cutoffs (\ref{eq_DR4}) can only be approximate. Indeed, this derivation assumed that fluctuations (\ref{eq_LD1b}) with different $h$ are independent, which is unlikely the case. Thus, we expect that the gradual breaking of hidden symmetry affects all structure functions at all scales of the intermittent dissipation range (\ref{eq_DR4b}).   
This point of view is confirmed in Fig.~\ref{fig3}. One can see that viscous effects change the pre-factors $C_p$ of power laws (\ref{eq1_SpH11}), which deviate from constant values in the range (\ref{eq_DR4b}). The cutoff (\ref{eq_DR4}) marks a scale, at which the deviation becomes so large that it overcomes the power-law dependence.

In summary, we established a precise definition of the inertial interval as the range of scales (\ref{eq_IRdef}), at which the hidden symmetry is restored in a statistical sense. Breaking of hidden symmetry in the forcing range does not depend on the (very large) Reynolds number, and the convergence to a hidden-symmetric state is exponential in shell number.
Breaking of hidden symmetry at small scales occurs gradually in the intermittent dissipation range (\ref{eq_DR4b}), and it is controlled by the intermittently fluctuating local Reynolds numbers (\ref{eq_DR1L}). Despite the power laws for structure functions seem to extend beyond the inertial interval till the order-dependent cutoffs (\ref{eq_DR4}), viscous terms alter their pre-factors considerably. The latter has the practical implications: exponents $\zeta_p$ are estimated more accurately in the $p$-independent range of scales (\ref{eq_IRdef}), rather than in the larger interval extended till the cuf-off (\ref{eq_DR4}); see Fig.~\ref{fig2}. Though our analysis reveals that the complexity of dissipation range is closely related to the hidden self-similarity, a detailed analysis of dissipation scales remains beyond our current approach. We show, however, in the next section that the dissipation range can be understood for a different dissipation model.

\section{Hidden-symmetric dissipation}
\label{sec_diss}

We have shown in Sections \ref{sec_intII} and \ref{sec_visc} that stationary statistics at very large Reynold numbers is controlled by the hidden scaling symmetry of the rescaled (projected) formulation. The hidden self-similarity is restored in the inertial range (\ref{eq_IRdef}) and gets broken in the forcing and dissipation ranges. The transition mechanism from the inertial interval to the dissipation range is intricate, since local Reynolds numbers appearing in the rescaled formulation are intermittent. In this section, we describe a class of dissipative modifications of a shell model, whose rescaled formulations are not intermittent. In these systems, the hidden scale invariance can be generalized in a way, which encompasses all small (both inertial and dissipative) scales. Such extended self-similarity yields a deeper insight into the small-scale dynamics of developed turbulence and has potential practical applications, e.g., the development of effective dissipative closures. Also, this self-similarity provides a close analogue of the Kolmogorov's (K41) theory, in which a broken scaling symmetry (\ref{eq_S1}) with $h = 1/3$ is replaced with the restored hidden scale invariance.

\subsection{Model with a viscous cutoff}

We start with a specific model, which is constructed by modifying viscous terms in system (\ref{eq1a}).
First, we define a sharp cutoff at the scale $\ell_s$ by setting 
    \begin{equation}
	u_n \equiv 0\ \ \textrm{for}\ \ n  > s. 
	\label{eq_HD0}
    \end{equation}
For $n \le 0$, we keep the same boundary conditions (\ref{eq2b}).
For the remaining shells, we consider the equations
    \begin{equation}
    \frac{du_n}{dt} = 
    k_0 \mathcal{B}^{(s)}_n[u], \quad n = 1,\ldots,s,
    \label{eq_HD1d}
    \end{equation}
where we introduce quadratic terms $\mathcal{B}^{(s)}_n[u]$ extending the definition (\ref{eq1Bn}) as 
    \begin{equation}
    \mathcal{B}^{(s)}_n[u] = \mathcal{B}_n[u]+\left\{
    \begin{array}{ll}
    -2^{n}|u_n|u_n, & n = s-1,s; \\[5pt]
    0,& \textrm{otherwise}.
    \end{array}
    \right.
    \label{eq_RE3}
    \end{equation}
In this new model, the dissipative term $-2^n|u_n|u_n$ is added in the equations of last two shells, $s-1$ and $s$. Remaining equations for $n = 1,\ldots,s-2$ have the ideal (inviscid) form (\ref{eq1Euler}).

One can verify the energy balance equation 
	\begin{equation}
	\frac{d\mathcal{E}}{dt} = \Pi_1-k_{s-1}|u_{s-1}|^3-k_s|u_s|^3,
    	\label{eq_HD3e}
    	\end{equation}
where $\mathcal{E}[u] = \frac{1}{2}\sum |u_n|^2$ is the total energy. Here $\Pi_1[u] = k_1\,\mathrm{Im}\left(
u_2u_1^*u_0\right)$ is the flux of energy from shell $n = 0$ to shell $1$, which is produced by the boundary condition (\ref{eq2b}); see Eqs.~(\ref{eq_EF2}) and (\ref{eq_EF1}). The terms $-k_{s-1}|u_{s-1}|^3-k_s|u_s|^3$ describe the energy dissipation at the cutoff shells $s-1$ and $s$. 

One can check that Eqs.~(\ref{eq_HD1d})--(\ref{eq_RE3}) of our new model are invariant with respect to the space-time scalings
    \begin{equation}
    t,\ u_n,\ s \ \ \mapsto \ \ 2^{1-h}t,\ 2^h u_{n+1},\ s-1
    \label{eq_S1new}
    \end{equation}
for any $h \in \mathbb{R}$.
This relation is analogous to the symmetry (\ref{eq_S1}), but with the cutoff shift $s \mapsto s-1$ substituting the scaling of viscosity. Symmetry (\ref{eq_S1new}) is broken at the integral scale $\ell_0$ by boundary conditions (\ref{eq2b}).

\subsection{Extended hidden scale invariance}
\label{subsec_HSbbb}

Let us transform Eqs.~(\ref{eq_HD1d})--(\ref{eq_RE3}) using rescaled variables (\ref{eq2}) with $n = m+N$.
Similarly to Section~\ref{subsec_RE} and Appendix~\ref{secA1}, in the interval $0 < m+N \le s$, one derives
    \begin{equation}
    \frac{dU_N^{(m)}}{d\tau^{(m)}} 
    = \mathcal{B}_N^{(S)}[U^{(m)}]
         -U_N^{(m)}\sum_{j = 0}^{m-1}\alpha^j\mathrm{Re}\left(U_{-j}^{(m)*}\mathcal{B}_{-j}^{(S)}[U^{(m)}]\right),
         \quad -m < N \le S,
    \label{eq_RE2}
    \end{equation}
where we introduced 
    \begin{equation}
	S = s-m.
    \label{eq_RE2_S}
    \end{equation}
The cutoff (\ref{eq_HD0}) implies that
    \begin{equation}
	U_N^{(m)} \equiv 0,\quad N > S.
    \label{eq_RE1}
    \end{equation}
At boundary shells, $N \le -m$, the same conditions (\ref{eq1R_bc}) remain valid.

Let us consider arbitrary small scales $\ell_m$ and $\ell_{m+N}$, which satisfy the conditions
    \begin{equation}
     \ell_s \le \ell_m \sim \ell_{m+N} \ll \ell_0.
    \label{eq_HD1}
    \end{equation}
Then, as in Section~\ref{subsec_HS}, one can neglect the upper bound in the exponentially converging sum of Eq.~(\ref{eq_RE2}). This yields
    \begin{equation}
    \frac{dU_N}{d\tau} 
    = \mathcal{B}_N^{(S)}[U]
         -U_N\sum_{j \ge 0}\alpha^j\mathrm{Re}\left(U_{-j}^{*}\mathcal{B}_{-j}^{(S)}[U]\right),
    \label{eq_RE2ss}
    \end{equation}
where we also dropped the superscript $(m)$ in order to simplify the introduction of hidden symmetry. 
One can verify that Eq.~(\ref{eq_RE2ss}) is invariant under the transformation
    \begin{equation}
	U_N,\ d\tau,\ S \ \mapsto \ \frac{U_{N+1}}{\sqrt{\alpha+|U_1|^2}},\ 2\sqrt{\alpha+|U_1|^2}\, d\tau,\ S.
	\label{eq_NM1}
    \end{equation}
Derivation of this invariance follows the same steps as in Appendix~\ref{secA_2}. 
It corresponds to the change of the reference shell $m \mapsto m+1$ simultaneously with the cutoff shell $s \mapsto s+1$, hence, leaving their difference $S = s-m$ intact.
We refer to the transformation (\ref{eq_NM1}) as the \textit{extended hidden symmetry}. This symmetry is similar to (\ref{eqI_HS})--(\ref{eqI_HSexp}), but it takes into account the dissipative terms and the cutoff in our new model. Thus, though the extended hidden symmetry is still broken at large scales due to boundary conditions, it is not broken at small dissipation scales.

Formulation of symmetry (\ref{eq_NM1}) does not depend on the exponent $h$, which makes it a weaker symmetry: it can be restored in a statistical sense even if all original scaling symmetries (\ref{eq_S1new}) are broken. Notice that different cutoff shells $s$ define different systems of equations. Thus, transformation (\ref{eq_NM1}) relates solutions of different models with different $s$. This is different from the hidden symmetry (\ref{eqI_HS})--(\ref{eqI_HSexp}) of the ideal system, which relates solutions for the same system.

\subsection{Small-scale statistics of multipliers}
\label{subsec_HSbb}

\begin{figure}[tp]
\centering
\includegraphics[width=0.8\textwidth]{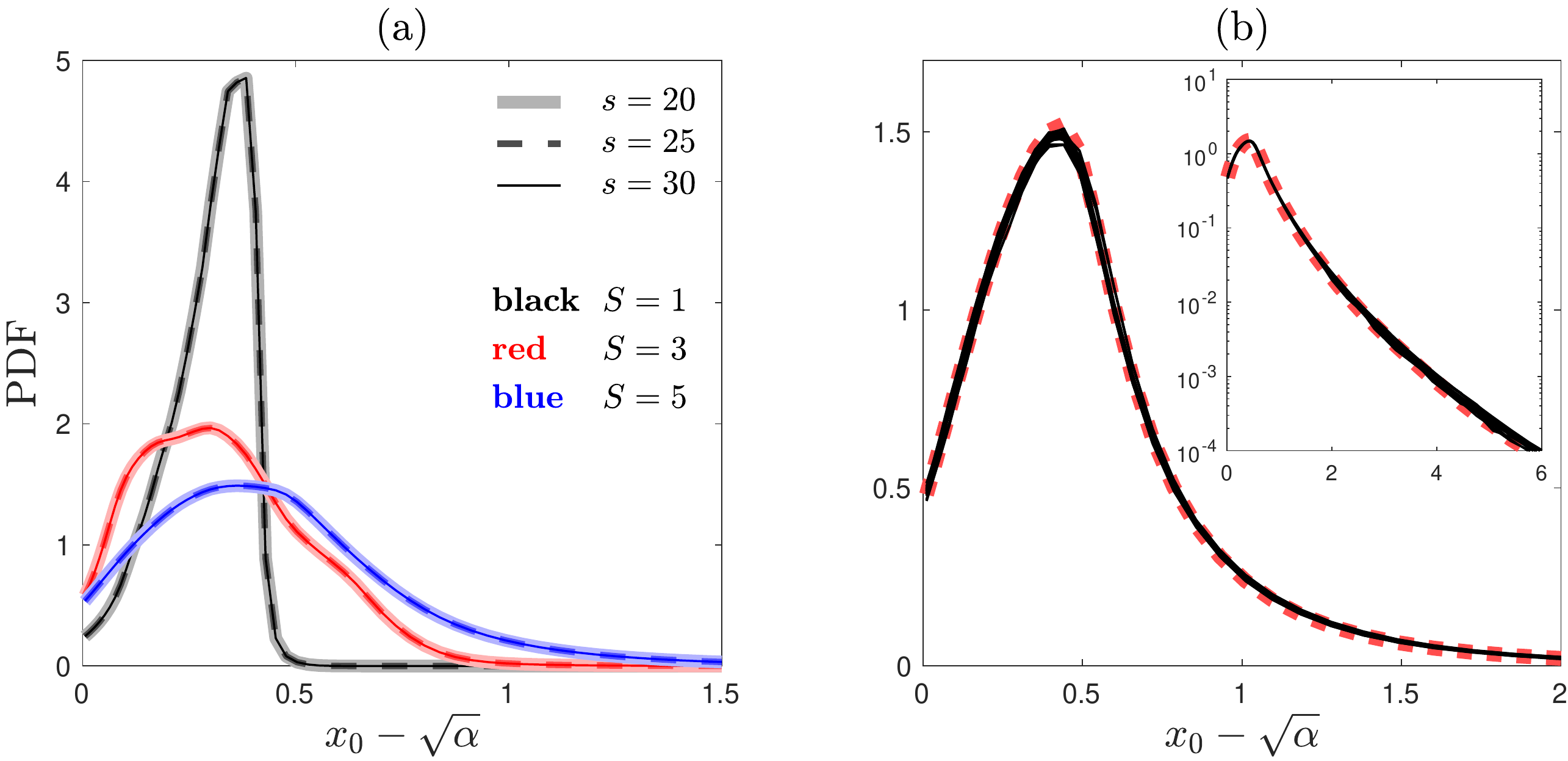}
\caption{(a) Verification of the extended hidden symmetry: PDFs of multipliers $x_0 = \mathcal{X}_0[U^{(m)}]$ depend only on the difference $S = s-m$. Shown are three graphs with the cutoff $s = 20,25,30$ for each $S = 1,3,5$. (b) PDFs of multipliers for $s = 30$, $m = 6,\ldots,15$ and $s = 25$, $m = 6,\ldots,10$ collapsing onto the inertial-interval PDF (dotted red line) from Fig.~\ref{fig1}(a). The inset presents the same plot with a vertical logarithmic scale.}
\label{fig7}
\end{figure}

We now verify numerically that the extended hidden symmetry (\ref{eq_NM1}) is restored in small-scale statistics of our viscous cutoff model. We will use the multiplier $x_0 = \mathcal{X}_0[U^{(m)}]$ depending on time $\tau^{(m)}$ as the observable. The symmetry (\ref{eq_NM1}) implies that the PDF of this multiplier depends on shell numbers $s$ and $m$ through their difference $S = s-m$ only. This property is verified in Fig.~\ref{fig7}(a) for $S = 1,3,5$. As $S$ increases, the effect of viscous cutoff $s$ at shell $m$ decreases, and PDFs converge to their hidden-symmetric form in the inertial interval. This is confirmed in Fig.~\ref{fig7}(b). 

\begin{figure}[tp]
\centering
\includegraphics[width=0.77\textwidth]{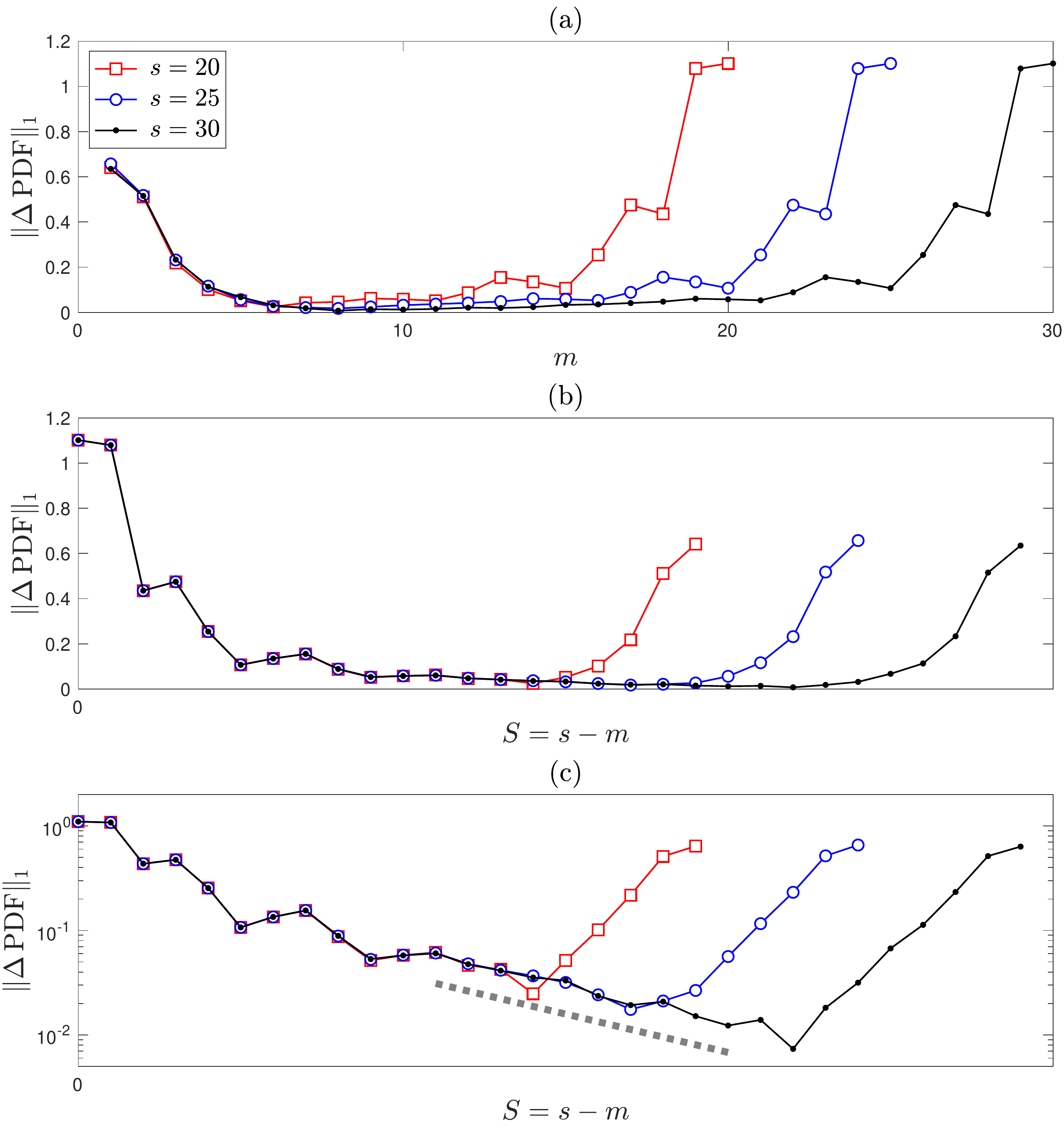}
\caption{(a) Difference between the PDF of multiplier $x_0 = \mathcal{X}_0[U^{(m)}]$ at shell $m$ and the hidden-symmetric PDF from the middle of the inertial interval. The results are shown for cutoff shells $s = 20,25,30$. (b,c) Same plots as functions of $S = s-m$ are presented in vertical linear and logarithmic scales. They collapse both in the inertial interval and the dissipation region, confirming the extended hidden symmetry. The dotted line demonstrates an exponential convergence to the inertial-interval PDF at large $S$.}
\label{fig8}
\end{figure}

Let us study the dependence of statistical distributions on $S = s-m$. Figure~\ref{fig8} shows a difference (measured with $L^1$-norm) between the multiplier PDF for any $m$ and the hidden-symmetric PDF from the inertial interval. The results are shown for different cutoffs, $s = 20,25,30$, as functions of the shell number $m$ (panel a) and  $S$ (panels b and c). Figures~\ref{fig8}(b,c) shown in linear and logarithmic vertical scales demonstrate a precise self-similarity at all small scales, i.e., both in the inertial interval and dissipation range. One can also see that the decay to the inertial-interval PDF has the exponential asymptotic form 
    \begin{equation}
    \label{eq_LexpD}
    \|\Delta\,PDF\|_1 \propto 2^{-\zeta_D S}, \quad
    \zeta_D \approx 0.25.
    \end{equation}
as shown by the dotted line in Fig.~\ref{fig8}(c). 

Symmetry (\ref{eq_NM1}) is broken at large scales of the forcing range. In the forcing range, the statistics coincides with the one of the original viscous shell model as described in Section~\ref{subsec_FR} and verified in Figs.~\ref{fig5b} and \ref{fig5c}. This reflect the independence of large-scale statistics not only of the cutoff scale, but also of a specific dissipation mechanism. 

In conclusion, stationary statistics in our viscous-cutoff model restores the extended hidden symmetry at small scales, both in the inertial interval and dissipation range. Far from the forcing and cutoff shells, the distribution approaches exponentially the universal distribution of the inertial interval. We interpret these exponential convergences on both sides of the inertial interval as the leading (left- and right-side) perturbation modes of a stable hidden-symmetric state described in Section~\ref{sec_intII}.

\subsection{Self-similarity of structure functions}
\label{subsec_EHS_mult}

\begin{figure}[pth]
\centering
\includegraphics[width=0.85\textwidth]{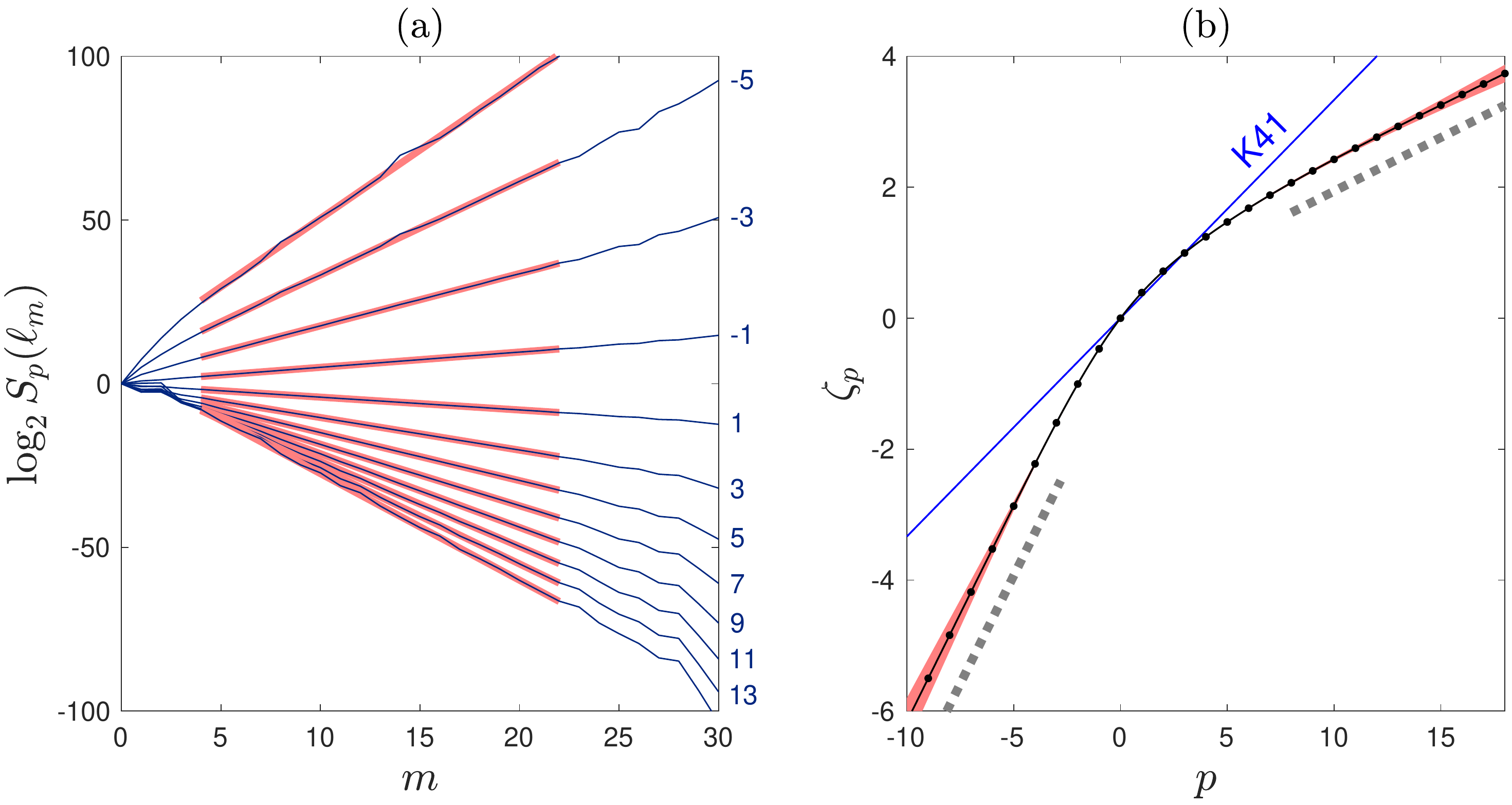}
\caption{(a) Logarithms of structure functions $S_p(\ell_m) = \left\langle \mathcal{A}_m^p[u] \right\rangle_t$ for odd orders $p = -7,-5,\ldots,15$ in the viscous cutoff model with $s = 30$. Bold red lines show the power law dependence $S_p(\ell_m) \propto \ell_m^{\zeta_p}$ in the inertial interval. (b) Anomalous exponents $\zeta_p$ computed in the interval $-10 \le p \le 18$ with the step $\Delta p = 0.2$. The shaded area indicates error bounds, and the blue line corresponds to the K41 linear dependence $p/3$. Dashed lines indicate the slopes $h_{\max}$ and $h_{\min}$ at large negative and positive $p$. The results use simulations with $T = 2\times 10^4$.}
\label{fig9}
\end{figure}

\begin{figure}[pth]
\centering
\includegraphics[width=0.85\textwidth]{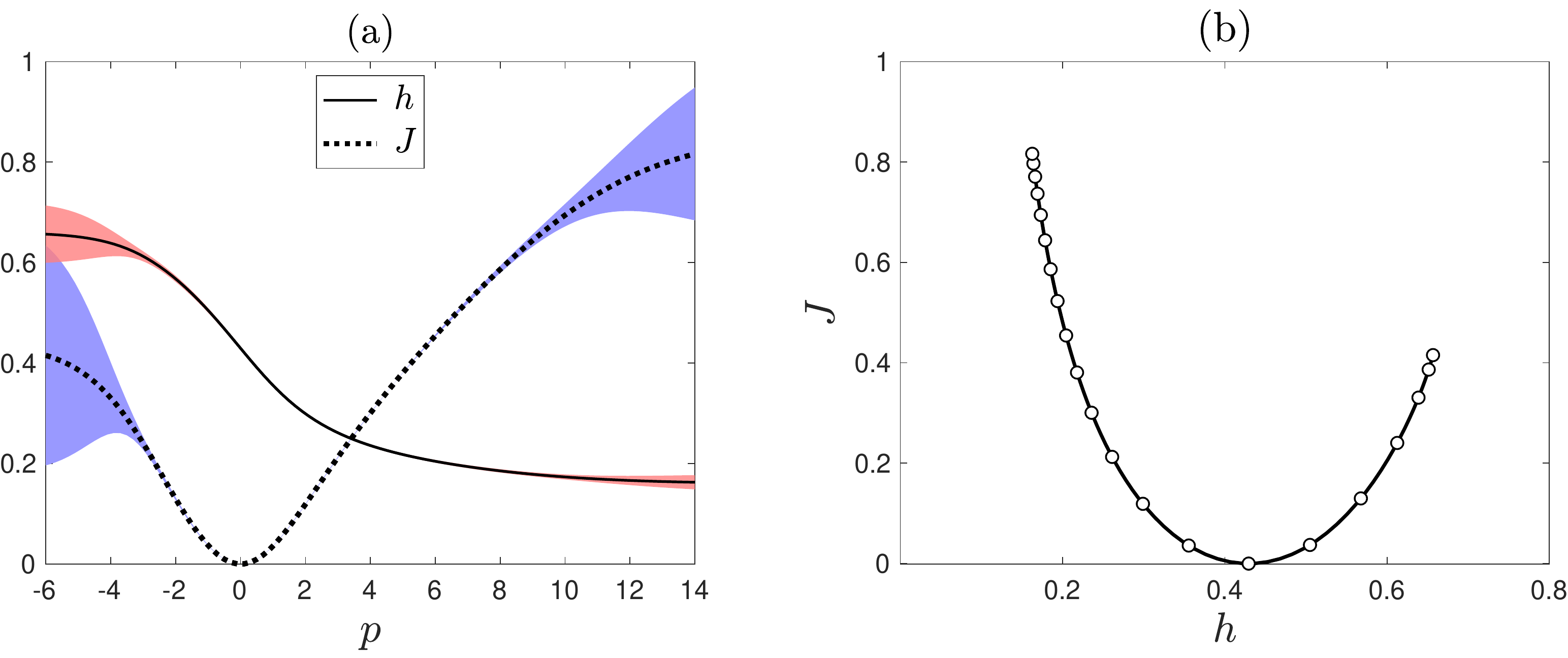}
\caption{(a) Scaling exponent $h$ and fractal codimension $J$ given by Eq.~(\ref{eq_LD1bJimp}) as functions of $p$ for the viscous cutoff model. The shaded area indicates error bounds. (b) The resulting function $J(h)$. Circles correspond to integer values of $p = -6,\ldots,14$. The results use simulations with $s = 30$ and $T = 2 \times 10^4$.}
\label{fig10}
\end{figure}

We now return to original variables and analyze the scaling of structure functions $S_p(\ell_m) = \left\langle \mathcal{A}_m^p[u] \right\rangle_t$ at small scales. The formulation and conclusions based on the hidden scale invariance (\ref{eqI_HS})--(\ref{eqI_HSexp}) in Section~\ref{sec_intII} remain valid in the inertial interval of the viscous cutoff model. We recall the general formulas (\ref{eq1_SpM}), (\ref{eq1_SpMM}) and (\ref{eq1_SpM5}) expressing structure functions in terms rescaled variables, which yield the asymptotic power-law scaling (\ref{eq1_SpH11}) as a consequence of the hidden scale invariance (\ref{eq1_SpH9a}) with Perron--Frobenius eigenmode (\ref{eq1_SpH9}). Figure~\ref{fig9} verifies the power law scaling with anomalous exponents $\zeta_p$, which agree within error bounds with those of the original viscous model in Fig.~\ref{fig2}. Figure~\ref{fig10} presents the fractal codimension $J(h)$ as described in Section~\ref{subsec_ldt}, in agreement with Fig.~\ref{fig2h} of the original shell model.


Extended hidden symmetry (\ref{eq_NM1}) is not limited to the inertial interval, and it relates statistical properties of rescaled variables for different cutoffs $s$. Let us show that this symmetry implies scaling relations for structure functions, which extend to the dissipation range. For this purpose, we introduce a sufficiently large integer $d$ estimating a number of shells in the dissipation range. 
This number $d$ does not depend on the cutoff shell $s$, as follows from the extended hidden symmetry. 
Then, the dissipation range contains the scales $\ell_m$ satisfying the inequalities
	\begin{equation}
	\ell_s \le \ell_m \lesssim \ell_{s-d},
    	\label{eq_HSC1}
    	\end{equation}
and the remaining small scales 
	\begin{equation}
	\ell_{s-d} \lesssim \ell_m \ll \ell_0 
    	\label{eq_HSC1_ii}
    	\end{equation}
belong to the inertial interval.

At scales of the inertial interval, expression (\ref{eq1_SpH10}) yields
	\begin{equation}
	d\mu_p^{(m)} \approx C_p \lambda_p^m \, d\nu_p. 
    	\label{eq_HSC1M}
    	\end{equation}
This relation is not valid for smaller scales of the dissipation range, where the hidden symmetry (\ref{eqI_HS})--(\ref{eqI_HSexp}) is broken. However, the extended version of the hidden symmetry (\ref{eq_NM1}) remains valid. It implies that the statistical properties depend on $m$ and $s$ only through their difference $S = s-m$. Hence, in the dissipation range (\ref{eq_HSC1}), we can replace Eq.~(\ref{eq1_SpH9a}) by the relation
	\begin{equation}
	p^{(m)}(x_1|\mathbf{x}_\ominus) \approx 
	\rho^{(S)}(x_1|\mathbf{x}_\ominus), 
	\ \ \mathcal{L}_p^{(m)} \approx \Lambda_p^{(S)},
	\label{eq_HSC2}
	\end{equation}
where we introduced the conditional probability density $\rho^{(S)}$ indexed by $S = s-m$, and the linear operator $\Lambda_p^{(S)}$ is given by Eq.~(\ref{eq1_SpM4inf}) with the density $\rho^{(S)}$. Consider a general relation (\ref{eq1_SpM5}) in the dissipation range (\ref{eq_HSC1}). Using Eq.~(\ref{eq_HSC2}) and then Eq.~(\ref{eq_HSC1M}) at scale $s-d$, we obtain 
    \begin{equation}
    	\begin{array}{rcl}
	    d\mu_p^{(m)} 
	    & = & 
	    \mathcal{L}_p^{(m-1)} \circ \mathcal{L}_p^{(m-2)} \circ \cdots \circ \mathcal{L}_p^{(s-d)}[d\mu_p^{(s-d)}] \\[3pt]
	    & \approx & 
	    C_p \lambda_p^{s-d}
	    \Lambda_p^{(S+1)} \circ \Lambda_p^{(S+2)} \circ \cdots \circ \Lambda_p^{(d)}[d\nu_p] \\[3pt]
	    & = & 
	    C_p 2^{-\zeta_p m} 2^{-\zeta_p (S-d)}
	    \Lambda_p^{(S+1)} \circ \Lambda_p^{(S+2)} \circ \cdots \circ \Lambda_p^{(d)}[d\nu_p],
	\end{array}
    \label{eq1_SpM5x}
    \end{equation}
where $\zeta_p = -\log_2 \lambda_p$.
Substituting (\ref{eq1_SpM5x}) into Eq.~(\ref{eq1_SpM}) and recalling that $2^{-m} = \ell_m/\ell_0$ and $2^S = \ell_m/\ell_s$ yields the final expression
	\begin{equation}
	S_p(\ell_m) \approx C_p u_0^p \left(\frac{\ell_m}{\ell_0}\right)^{\zeta_p}\,F_p\left(\frac{\ell_m}{\ell_s}\right),\quad 
	\zeta_p = -\log_2 \lambda_p,
    	\label{eqEHS_5}
    	\end{equation}
with the function 
	\begin{equation}
	F_p\left(2^S\right) = \left\{
	\begin{array}{ll}
		2^{-\zeta_p (S-d)} \int \Lambda_p^{(S+1)} \circ 
		\Lambda_p^{(S+2)} \circ \cdots \circ 
		\Lambda_p^{(d)}[d\nu_p], & 0 \le S < d; \\[3pt]
		1,& S \ge d.
	\end{array}\right.
    	\label{eqEHS_6}
    	\end{equation}
Here we also assigned the value $F_p\left(2^S\right) = 1$ for $S \ge d$, such that Eq.~(\ref{eqEHS_5}) recovers the power law (\ref{eq1_SpH11}) in the inertial interval. 
Thus, expression (\ref{eqEHS_5}) describes the universal form of structure functions at all small scales up to the cutoff.
	
One can recognize some analogy of relation (\ref{eqEHS_5}) with the formula from Kolmogorov's K41 theory~\cite{kolmogorov1941local,frisch1999turbulence}, which describes a functional dependence of energy spectrum on a wavenumber and a viscous micro-scale. However, the K41 theory uses the scaling symmetry (\ref{eq_S1}) with $h = 1/3$, which is broken in the turbulent statistics. On the contrary, relation (\ref{eqEHS_5}) follows from the extended hidden symmetry (\ref{eq_NM1}), which is restored in a statistical sense. 

\begin{figure}[t]
\centering
\includegraphics[width=0.95\textwidth]{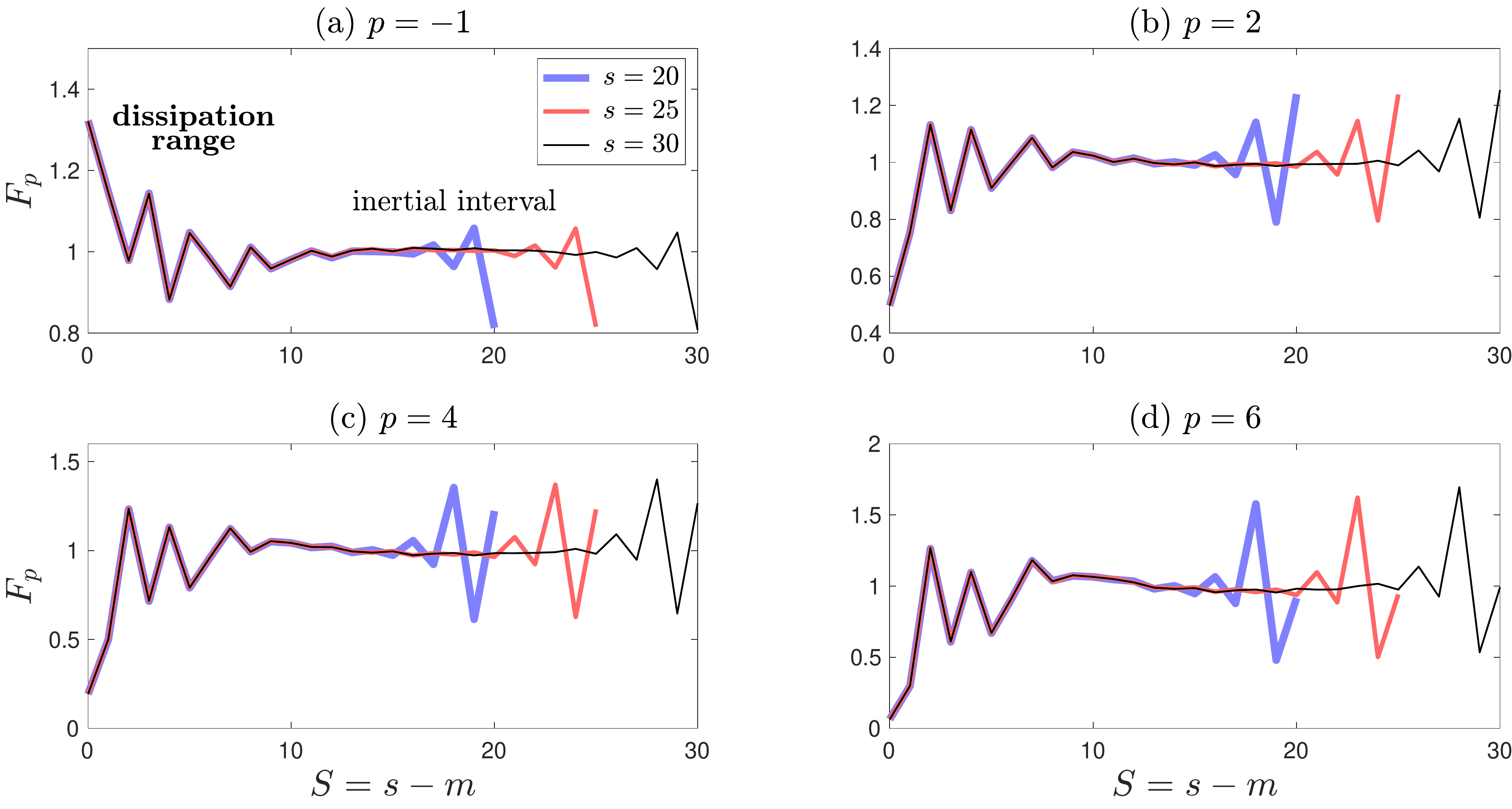}
\caption{Compensated structure functions (\ref{eqEHS_8}) of orders $p = -1,2,4,6$ for cutoffs $s = 20,25,30$. Using $S = s-m$ in the horizontal axis, these functions collapse both in the dissipation range and inertial interval.}
\label{fig11}
\end{figure}

We verify relation (\ref{eqEHS_5}) with numerical simulations in Fig.~\ref{fig11}. Here, using simulations with $s = 20,25,30$, we plot the compensated structure functions 
	\begin{equation}
	\frac{1}{C_p u_0^p}\left(\frac{\ell_m}{\ell_0}\right)^{-\zeta_p}\, S_p(\ell_m) \approx \,F_p\left(\frac{\ell_m}{\ell_s}\right)
    	\label{eqEHS_8}
    	\end{equation}
as functions of $s-m = \log_2(\ell_m/\ell_s)$. In agreement with (\ref{eqEHS_5}) one observes a high-quality collapse of these functions both in the dissipation range and inertial interval. The collapse gets broken only in the forcing range, which is located on the right side of each graph.

\subsection{General form of hidden-symmetric dissipation}

The model (\ref{eq_HD0})--(\ref{eq_RE3}) is just a specific example in a large class of viscous-cutoff models possessing the extended hidden symmetry (\ref{eq_NM1}). One can use any dissipative terms in $\mathcal{B}_n^{(s)}[u]$, provided that they have the same scaling properties as Eq.~(\ref{eq_RE3}). Precisely, this condition is formulated as the positive homogeneity 
    \begin{equation}
    \mathcal{B}_n^{(s)}[a u] = a^{2}\mathcal{B}_n^{(s)}[u], \quad a > 0,
    \label{eq_GC1}
    \end{equation}
and the scaling relation
    \begin{equation}
    \mathcal{B}_n^{(s)}[u] = 2\mathcal{B}_{n-1}^{(s-1)}[u'], \quad
    u' = (u'_{j})_{j \in \mathbb{Z}} = (u_{j+1})_{j \in \mathbb{Z}}.
    \label{eq_GC1b}
    \end{equation}
One can verify that derivations of Section~\ref{subsec_HSbbb} remain valid and lead to the rescaled formulation (\ref{eq_RE2})--(\ref{eq_RE1}) with the extended hidden symmetry (\ref{eq_NM1}). 

Recovery of the extended hidden symmetry can facilitate future theoretical studies. For example such models are more convenient for more detailed analysis of hidden-symmetric states. 
Also, these models provide a general framework for optimal closures, which are universal with respect to the cutoff scale; see \cite{biferale2017optimal,ortali2022towards,juliathesis} for recent studies in this direction. 

\section{Conclusion}\label{secD}

In this work we described how the hidden scaling symmetry governs the turbulent statistics in a shell model at all scales of motion. The hidden symmetry emerges when equations of motion are written for new velocities and time, which are rescaled dynamically. Geometrically, this rescaling is a projection in the phase space~\cite{mailybaev2020hidden}. Through this work, we assume that the hidden symmetry is recovered in a given statistical distribution, as numerical simulations strongly suggest. An important direction for future research would be the formulation of equations (in a spirit of~\cite{eyink2003gibbsian}) and stability analysis of the hidden-symmetric statistics.

The paper contains three parts. The first part is dedicated to the inertial interval only. Here we recalled the previous results~\cite{mailybaev2022shell} relating the anomalous scaling laws of structure functions with Perron--Frobenius eigenmodes induced by the hidden-symmetric statistics. We generalized these results to a larger class of observables. In particular, we described how a constant flux of energy to small scales (the dissipative anomaly~\cite{eyink2006onsager}) emerges in a hidden-symmetric framework. Also, we showed that the hidden self-similarity leads naturally to the large deviation theory, thereby providing a theoretical justification of the Parisi-Frisch multifractal approach from first principles.

The second part is devoted to breaking of the hidden symmetry by forcing and viscous dissipation. In the transition from the forcing range to the inertial interval, we observed an exponentially decaying mode of the stable hidden-symmetric state. The breaking of hidden symmetry in the dissipation region is more intricate because of intermittently fluctuating viscous terms in the rescaled formulation. We studied how this intermittency affects power laws and pre-factors of structure functions. Such intermittency is closely related to the concept of intermediate dissipation range introduced in \cite{frisch1993prediction}. 

In the third part, we demonstrated that the intermittency of rescaled viscous terms is not a general feature, but rather an artifact of the viscous dissipation model. We introduced a viscous-cutoff model with modified dissipation terms, for which the hidden symmetry extends to all scales of the dissipation range. This symmetry yields a self-similar functional form of structure functions valid through the whole range of small scales, i.e., both in the inertial interval and dissipation range. This result resembles the original Kolmogorov's self-similarity of the dissipation range, but formulated now for the statistically restored hidden symmetry.

The hidden symmetry has an analogous formulation for the incompressible Navier--Stokes system, as shown both within a general theoretical framework~\cite{mailybaev2020hidden} and by explicit derivations~\cite{mailybaev2022hidden}. The rescaled formulation of the Navier--Stokes system combines a projection induced by time scalings (used in the shell model) with a projection induced by the Galilean symmetry (equivalent to the quasi-Lagrangian description~\cite{belinicher1987scale,l1991scale}). 
Though the intermittency is removed from the inertial interval in the rescaled formulation, the rescaled viscous terms become intermittent~\cite{mailybaev2022hidden}, just like in the shell model. This comparison suggests that the approach developed here for the shell model is applicable for an analogous study of the 3D incompressible Navier--Stokes turbulence. In particular, modified viscous-cutoff models, which recover an extended hidden symmetry at all small scales, may help to overcome the current limitations of numerical simulations; we refer to a recent numerical study~\cite{biferale2019self} pursuing a similar goal from a different point of view.

\section{Appendix}\label{sec8}

\subsection{Derivation of rescaled equations (\ref{eq1Ra})--(\ref{eq1Rb})}
\label{secA1}

Dropping the superscripts $(m)$ for simplicity and using  Eqs.~(\ref{eq2A})--(\ref{eq2}), we write 
    \begin{equation}
    \begin{array}{rcl}
	    \displaystyle
	    \frac{dU_N}{d\tau} 
	    & = & \displaystyle
	    \mathcal{T}_m\frac{d}{dt} \frac{u_{m+N}}{\mathcal{A}_m}
	    = \frac{1}{k_m\mathcal{A}_m^2}\frac{du_{m+N}}{dt} 
	    -\frac{u_{m+N}}{2k_m\mathcal{A}_m^4}\frac{d\mathcal{A}_m^2}{dt} 
	    \\[15pt]
	    & = & \displaystyle
	    \frac{1}{k_m\mathcal{A}_m^2}\frac{du_{m+N}}{dt} 
	    -\frac{u_{m+N}}{k_m\mathcal{A}_m^4}
	    \sum_{j \ge 0}\alpha^{j} \,\mathrm{Re} \left(u_{m-j}^*\frac{du_{m-j}}{dt}\right)
	    \\[15pt]
	    & = & \displaystyle
	    \frac{1}{k_m\mathcal{A}_m^2}\frac{du_{m+N}}{dt} 
	    -\frac{U_N}{k_m\mathcal{A}_m^2}
	    \sum_{j \ge 0}\alpha^{j} \,\mathrm{Re} \left(U_{-j}^*\frac{du_{m-j}}{dt}\right).
    \end{array}
    \label{eqA1_1}
    \end{equation}
Using Eqs.~(\ref{eq2b})--(\ref{eq1a}) with $k_{m} = 2^m k_0$ and then Eqs.~(\ref{eq1Bn}), (\ref{eq2}) and (\ref{eq1Rb}),  yields
    \begin{equation}
    \frac{1}{k_m\mathcal{A}_m^2}\frac{du_{m+N}}{dt} 
    = \frac{\mathcal{B}_{m+N}[u]}{2^m\mathcal{A}_m^2}
	-\frac{\nu k_m 4^N u_{m+N}}{\mathcal{A}_m^2}
	= \mathcal{B}_N[U]-\frac{\nu k_m}{\mathcal{A}_m}\, 4^N U_{N}
	= \mathcal{B}_N[U]-\frac{4^N U_{N}}{\mathrm{R}_m[u]}.
    \label{eqA1_2}
    \end{equation}
The similar expression with $N \mapsto -j$ for $j \le m-1$ reads
    \begin{equation}
    \frac{1}{k_m\mathcal{A}_m^2}\frac{du_{m-j}}{dt} 
	= \mathcal{B}_{-j}[U]-\frac{4^{-j} U_{-j}}{\mathrm{R}_m[u]},
    \label{eqA1_2j}
    \end{equation}
and the derivatives with $j \ge m$ vanish by boundary conditions (\ref{eq2b}).
Using expressions (\ref{eqA1_2}) and (\ref{eqA1_2j}) in Eq.~(\ref{eqA1_1}) yields the rescaled equation~(\ref{eq1Ra}).

It immediately follows from boundary conditions (\ref{eq2b}) and relation (\ref{eq2}) that $U_N^{(m)} = 0$ for $N < -m$. Using conditions (\ref{eq2b}) and squared first relation in (\ref{eq2A}), one can express
    \begin{equation}
	u_0 = \sqrt{\alpha^{-m}\mathcal{A}_m^2[u]-\sum_{j = 0}^{m-1}\alpha^{j-m}|u_{m-j}|^2}.
        \label{eqA1_3n}
    \end{equation}
Then, relations (\ref{eq2}) and (\ref{eqA1_3n}) for the remaining rescaled velocity $U_{-m}^{(m)}$ yield
    \begin{equation}
    U_{-m}^{(m)} = \frac{u_0}{\mathcal{A}_m[u]}
    = \sqrt{\alpha^{-m}-\sum_{j = 0}^{m-1} \alpha^{j-m}\left(\frac{|u_{m-j}|}{\mathcal{A}_m[u]}\right)^2}
    = \sqrt{\alpha^{-m}-\sum_{j = 0}^{m-1} \alpha^{j-m}\big|U_{-j}^{(m)}\big|^2}.
        \label{eqA1_4n}
    \end{equation}

\subsection{Derivation of the hidden scaling symmetry}
\label{secA_2}

We now show the invariance of system (\ref{eqI_R}) with respect to the hidden scaling transformation (\ref{eqI_HS})--(\ref{eqI_HSexp}).
Using the new variables given by (\ref{eqI_HSexp}), we write
	\begin{equation}
	\label{eqA1}
	\frac{d\hat{U}_N}{d\hat{\tau}}
	= \frac{1}{2\sqrt{\alpha+|U_1|^2}} \,
	\frac{d}{d\tau} \frac{U_{N+1}}{\sqrt{\alpha+|U_1|^2}}
	= \frac{1}{2(\alpha+|U_1|^2)} \,
	\frac{d U_{N+1}}{d\tau}
	-\frac{U_{N+1}}{2(\alpha+|U_1|^2)^2}\,\mathrm{Re}\left(U_1^*\frac{dU_1}{d\tau}\right).
	\end{equation}
Substituting the derivatives from (\ref{eqI_R}) yields
	\begin{equation}
	\label{eqA2}
	\begin{array}{rcl}
	\displaystyle
	\frac{d\hat{U}_N}{d\hat{\tau}}
	&=& 
	\displaystyle
	\frac{1}{2(\alpha+|U_1|^2)} \,
	\bigg(\mathcal{B}_{N+1}[U]
         -U_{N+1}\sum_{j \ge 0} \alpha^{j}\mathrm{Re}\left(U_{-j}^{*}\mathcal{B}_{-j}[U]\right)\bigg)
	\\[15pt] && \displaystyle
	-\frac{U_{N+1}}{2(\alpha+|U_1|^2)^2}\bigg(\mathrm{Re}\,(U_1^*\mathcal{B}_1[U])-|U_1|^2\sum_{j \ge 0} 
	\alpha^j \mathrm{Re}(U_{-j}^*\mathcal{B}_{-j}[U])\bigg)
	\\[17pt] & = & \displaystyle
	\frac{\mathcal{B}_{N+1}[U]}{2(\alpha+|U_1|^2)} 
	-\frac{U_{N+1}}{2(\alpha+|U_1|^2)^2}\sum_{j \ge -1} \alpha^{j+1}\mathrm{Re}\left(U_{-j}^{*}\mathcal{B}_{-j}[U]\right).
	\end{array}
	\end{equation}
Using Eqs.~(\ref{eqI_HSexp}) and (\ref{eq1Bn}) and changing the summation variable $j' = j+1$ yields
	\begin{equation}
	\label{eqA2nxt2}
	\frac{d\hat{U}_N}{d\hat{\tau}}
         = {\mathcal{B}}_{N}[\hat U]
	-\hat{U}_{N}\sum_{j' \ge 0} \alpha^{j}\mathrm{Re}\left(\hat{U}_{-j'}^{*}{\mathcal{B}}_{-j'}[\hat U]\right),
	\end{equation}
which has the same form as the original Eq.~(\ref{eqI_R}).

\subsection{Expressions for generalized multipliers}
\label{subsec_GMult}

Using Eq.~(\ref{eq2A}), we derive
    \begin{equation}
	\frac{\mathcal{A}_{n}[u]}{\mathcal{A}_{n-1}[u]} 
	= \sqrt{\frac{\sum_{j \ge 0}\alpha^j|u_{n-j}|^2}{\sum_{j \ge 0}\alpha^j|u_{n-1-j}|^2}}
	= \sqrt{\frac{|u_n|^2+\sum_{j \ge 1}\alpha^j|u_{n-j}|^2}{\sum_{j \ge 1}\alpha^{j-1}|u_{n-j}|^2}}
	= \sqrt{\alpha+\frac{\alpha|u_{n}|^2}{\sum_{j \ge 1}\alpha^j|u_{n-j}|^2}}.
	\label{eqKm_A}
    \end{equation}
Dividing both sides of the last fraction by $\mathcal{A}_m^2[u]$ and using Eq.~(\ref{eq2}) yields
    \begin{equation}
	\frac{\mathcal{A}_{n}[u]}{\mathcal{A}_{n-1}[u]} 
	= \sqrt{\alpha+\frac{\alpha|u_{n}|^2/\mathcal{A}_m^2[u]}{\sum_{j \ge 1}\alpha^j|u_{n-j}|^2/\mathcal{A}_m^2[u]}}
	= \sqrt{\alpha+\frac{\alpha \big|U_N^{(m)}\big|^2}{\sum_{j \ge 1}{\alpha^{j}\big|U_{N-j}^{(m)}\big|^2}}}
	= \mathcal{X}_N[U^{(m)}],
	\label{eqKm_A2}
    \end{equation}
where $n = m+N$ and the last expression follows from definition (\ref{eqKm2}).

\subsection{Details of numerical simulations}
\label{subsecA_num}

In the shell model we take $\ell_0 = u_0 = 1$ and consider rescaled variables for $\alpha = 1/8$. Simulations are performed using MATLAB solvers~\cite{shampine1997matlab} 
(ode15s for the original viscous model and ode45 for the viscous-cutoff model), 
with the tolerances $RelTol = 10^{-7}$ and $AbsTol = 10^{-8}$. For initial conditions, we use the K41-like state $u_n = e^{-i\theta_n}k_n^{-1/3}$ with independent random phases $\theta_n$. Skipping the initial time interval $\Delta t = 20$, which contains transient behaviors, we use the data obtained in the long time interval $t \in [0,T]$. In simulations, we evaluate the rescaled times for different reference shells by integrating respective equations, which are written using definitions (\ref{eq2A})--(\ref{eq2}) as $d\tau^{(m)}/dt = k_m\mathcal{A}_m[u]$. 

We evaluate statistical properties using a standard histogram approach with weight coefficients proportional to time steps. Notice that statistics of rescaled variables must be computed using the respective time $\tau^{(m)}$. Computation of marginal densities $f_p(x_0)$ for measures $d\mu_p^{(m)}(\mathbf{x}_\ominus)$ is performed similarly with the extra weight coefficient, which follows from expression (\ref{eq1_SpMM}). An alternative way follows from the comparison of Eqs.~(\ref{eq1_Sp}) and (\ref{eq1_SpM}): one can estimate $f_p(x_0)$ as a sum of quantities $\mathcal{A}_m^p[u]\Delta t/(T\, \Delta x_0)$ within each bin of the multiplier $x_0$, where $\Delta t$ is a step in the original time and $\Delta x_0$ is a bin size. A similar histogram approach can be applied for computing the densities $f_\psi(\psi)$ from (\ref{eq_EF1x}).

Exponents $\zeta_p$ in Figs.~\ref{fig2} and \ref{fig9} are computed using the least squares method. The errors are estimated as four root-mean-square deviations divided by the number of interpolating shells. We stress that such error estimates are not very reliable, because they refer to statistical fluctuations only, while deviations can also be caused by forcing and dissipation effects.

\subsection{Structure functions in terms of rescaled variables}
\label{secA_3}

We write Eq.~(\ref{eq1_Sp}) as
    \begin{equation}
    S_p(k_m) = \lim_{T \to \infty} \frac{1}{T}\int_0^T \mathcal{A}_m^p[u] dt
    = \lim_{T \to \infty} \frac{\int_0^T \mathcal{A}_m^p[u] dt}{\int_0^T dt}.
    \label{eqZ_D3a}
    \end{equation}
After the change of time $dt = d\tau^{(m)} / (k_m \mathcal{A}_m)$ from (\ref{eq2A})--(\ref{eq2}), we obtain
    \begin{equation}
    S_p(k_m) 
    = \lim_{T^{(m)} \to \infty} \frac{\int_0^{T^{(m)}} \mathcal{A}_m^{p-1}[u] d\tau^{(m)}}{\int_0^{T^{(m)}} \mathcal{A}_m^{-1}[u] d\tau^{(m)}}
    = \frac{\langle \mathcal{A}_m^{p-1}[u] \rangle_{\tau^{(m)}} }{\langle \mathcal{A}_m^{-1}[u] \rangle_{\tau^{(m)}}},
    \label{eqZ_D3b}
    \end{equation}
where the limit $\tau^{(m)} = T^{{(m)}}$ corresponds to $ t = T$, and $\mathcal{A}_m[u]$ must be expressed as a function of $\tau^{(m)}$. For the latter, we use Eq.~(\ref{eq1_Sp2b}), which yields Eq.~(\ref{eq1_SpF}).

By definition, the joint statistics of multipliers $x_1 = \mathcal{X}_1[U^{(m)}]$ and $\mathbf{x}_\ominus = \boldsymbol{\mathcal{X}}_\ominus[U^{(m)}]$ is given by the probability measure 
    \begin{equation}
    	    p^{(m)}(x_1|\mathbf{x}_\ominus) \,dx_1\,d\mu^{(m)}(\mathbf{x}_\ominus).
    \label{eqAp_M4}
    \end{equation}
Let us find the measure $d\mu^{(m+1)}(\tilde{\mathbf{x}}_\ominus)$ describing the statistics of multipliers $\tilde{\mathbf{x}}_\ominus = \boldsymbol{\mathcal{X}}_\ominus [U^{(m+1}]$ as functions of time $\tau^{(m+1)}$.
The times $\tau^{(m)}$ and $\tau^{(m+1)}$ are related by Eqs. (\ref{eq2A}), (\ref{eq2}) and (\ref{eqKm}) as
    \begin{equation}
	\frac{d\tau^{(m+1)}}{d\tau^{(m)}} = \frac{k_{m+1}\mathcal{A}_{m+1}[u]}{k_{m}\mathcal{A}_{m}[u]} = 2\mathcal{X}_1[U^{(m)}].
	\label{eqAp_M5}
    \end{equation}
Hence, the change of time from $\tau^{(m)}$ to $\tau^{(m+1)}$ introduces the density factor $x_1/\langle \mathcal{X}_1[U^{(m)}] \rangle_{\tau^{(m)}}$ in the probability measure; see e.g. \cite[\S10.3]{cornfeld2012ergodic}. 
Using Eq.~(\ref{eqKm}), one derives the identity 
    \begin{equation}
    \tilde{\mathbf{x}}_\ominus = (\tilde{x}_0,\tilde{\mathbf{x}}_-) = (x_1,\mathbf{x}_\ominus)
    \label{eqAp_M3bb}
    \end{equation}
for the multipliers $\tilde{\mathbf{x}}_\ominus = \boldsymbol{\mathcal{X}}_\ominus[U^{(m+1)}]$. Using this relation and the time factor in Eq.~(\ref{eqAp_M4}), we obtain
    \begin{equation}
	d\mu^{(m+1)}(\tilde{\mathbf{x}}_\ominus)
	=  \frac{x_1\,p^{(m)}(x_1|\mathbf{x}_\ominus)}{\langle \mathcal{X}_1[U^{(m)}] \rangle_{\tau^{(m)}}}\,
	dx_1\,d\mu^{(m)}(\mathbf{x}_\ominus)
	= \frac{\tilde{x}_0\,p^{(m)}(\tilde{x}_0|\tilde{\mathbf{x}}_-)}{\langle \mathcal{X}_1[U^{(m)}] \rangle_{\tau^{(m)}}}\,
	d\tilde{x}_0\,d\mu^{(m)}(\tilde{\mathbf{x}}_-).
	\label{eqAp_M6}
    \end{equation}
Using Eqs.~(\ref{eq1_SpMM}), (\ref{eqAp_M6}) and (\ref{eqAp_M3bb}), we derive the expression (\ref{eq1_SpM4}) as follows
    \begin{equation}
    	\begin{array}{rcl}
    	    d\mu_p^{(m+1)}(\tilde{\mathbf{x}}_\ominus)
	    & = &\displaystyle
	    \frac{\big(\prod_{j = 0}^{m}\tilde{x}_{-j}\big)^{p-1}}{\int
	    \big(\prod_{j = 0}^{m}\tilde{x}_{-j}\big)^{-1} d\mu^{(m+1)}}\,
	    d\mu^{(m+1)}(\tilde{\mathbf{x}}_\ominus)
	    \\[19pt]
	& = &\displaystyle
	    \frac{\tilde{x}_0^p\,p^{(m)}(\tilde{x}_0|\tilde{\mathbf{x}}_-)\,
	    \big(\prod_{j = 1}^{m}\tilde{x}_{-j}\big)^{p-1}}{\int
	    p^{(m)}(\tilde{x}_0|\tilde{\mathbf{x}}_-) \,d\tilde{x}_0\,
	   \big(\prod_{j = 1}^{m}\tilde{x}_{-j}\big)^{-1}
	   d\mu^{(m)}(\tilde{\mathbf{x}}_-)}
	    \,d\tilde{x}_0\,d\mu^{(m)}(\tilde{\mathbf{x}}_-) \\[19pt]
	& = &\displaystyle
	    \tilde{x}_0^p\,p^{(m)}(\tilde{x}_0|\tilde{\mathbf{x}}_-) \,d\tilde{x}_0\,
	    \frac{
	   \big(\prod_{j = 1}^{m}\tilde{x}_{-j}\big)^{p-1} 
	   }{\int
	   \big(\prod_{j = 1}^{m}\tilde{x}_{-j}\big)^{-1}\, d\mu^{(m)}(\tilde{\mathbf{x}}_-)}
	    \,d\mu^{(m)}(\tilde{\mathbf{x}}_-)
	    \\[15pt]
	& = &\displaystyle
	    \tilde{x}_0^p\,p^{(m)}(\tilde{x}_0|\tilde{\mathbf{x}}_-) \,d\tilde{x}_0\,
	    d\mu_p^{(m)}(\tilde{\mathbf{x}}_-),
	\end{array}	
    \label{eqAp_M7}
    \end{equation}
where we also used the full probability condition $\int p^{(m)}(\tilde{x}_0|\tilde{\mathbf{x}}_-) \,d\tilde{x}_0 \equiv 1$. 
 
Consider now the sequence $\Psi_m[u]$ and rescaled velocities from Eq.~(\ref{eq2}). Using relation (\ref{eq_GS1}) with $a = \mathcal{A}_m[u]$ and relation (\ref{eq_GS1Prime}) iteratively $m$ times, we have
    \begin{equation}
	\Psi_m[u] = \Psi_0[U^{(m)}]\, \mathcal{A}_m^p[u].
	\label{eqAp_M8}
    \end{equation}
Then, similarly to Eq.~(\ref{eq1_SpF}) derived in (\ref{eqZ_D3a})--(\ref{eqZ_D3b}), one obtains
    \begin{equation}
    \langle \Psi_m[u] \rangle_t
    = \frac{\big\langle \Psi_0[U^{(m)}] \big(\prod_{j = 0}^{m-1}\mathcal{X}_{-j}[U^{(m)}]\big)^{p-1} \big\rangle_{\tau^{(m)}}
    }{\big\langle \big(\prod_{j = 0}^{m-1}\mathcal{X}_{-j}[U^{(m)}]\big)^{-1} 
    \big\rangle_{\tau^{(m)}}}.
    \label{eqAp_M9}
    \end{equation}
Consider the measure 
    \begin{equation}
    	    \rho_f^{(m)}(\psi|\mathbf{x}_\ominus) \, d\psi\, d\mu_0^{(m)}(\mathbf{x}_\ominus)
    \label{eqAp_M10}
    \end{equation}
describing the joint statistics of the variable $\psi = \Psi_0[U^{(m)}]$ and the $\mathbf{x}_\ominus = \boldsymbol{\mathcal{X}}_\ominus[U^{(m)}]$.
Using this measure in Eq.~(\ref{eqAp_M9}) yields the relation (\ref{eq_GS2}), where the measure $d\mu_p^{(m)}(\mathbf{x}_\ominus)$ is defined in (\ref{eq1_SpMM}).

\subsection{Derivations using the G\"artner--Ellis Theorem}
\label{subsec_LDT}

In \cite[\S3.3.1]{touchette2009large}, the G\"artner--Ellis Theorem is formulated under the assumption that
    \begin{equation}
    \lambda(k) = \lim_{m \to \infty} \frac{1}{m} \log \langle e^{mkX_m} \rangle 
    \label{eqLDT_1}
    \end{equation}
exists and is differentiable for all $k \in \mathbb{R}$, where $X_m$ is a sequence of real random variables with positive integer indices $m$. Then $X_m$ satisfy the Large Deviation Principle expressed as 
    \begin{equation}
    P(X_m \in [x,x+dx]) \sim e^{-mI(x)} dx,
    \label{eqLDT_2}
    \end{equation}
with a rate function $I(x)$ given by
    \begin{equation}
    I(x) = \sup_{k \in \mathbb{R}}\big( kx-\lambda(k) \big).
    \label{eqLDT_3}
    \end{equation}
Expression (\ref{eqLDT_3}) is the Legendre transform and its well-known inverse reads 
    \begin{equation}
    \lambda(k) = \sup_{x \in \mathbb{R}}\big( kx-I(x) \big).
    \label{eqLDT_4}
    \end{equation}
One can verify that Eqs.~(\ref{eqLDT_1})--(\ref{eqLDT_4}) take the form (\ref{eq_LD1cX})--(\ref{eq_LD1bJ}) and (\ref{eq_LD1inv}) if one identifies
    \begin{equation}
	k = p ,\ \ 
	X_m = -W_m\log 2, \ \ 
	x = -h \log 2, \ \  
	\lambda(k) = - \zeta_p \log 2, \ \  
	I(x)  = J(h) \log 2. 
    \label{eqLDT_5}
    \end{equation}

\vspace{2mm}\noindent\textbf{Acknowledgments.} 
The authors is grateful to Berengere Dubrulle, Dmytro Bandak, Luca Biferale and Simon Thalabard for useful discussions.
This work was supported by CNPq grant 308721/2021-7 and FAPERJ grant E-26/201.054/2022. 

\bibliographystyle{plain}
\bibliography{refs}

\end{document}